\documentclass[onecolumn,11pt,a4paper]{IEEEtran}
\usepackage{amsmath,amssymb,amsfonts}
\usepackage{algorithmic}
\usepackage{graphicx}
\usepackage{textcomp}
\usepackage{hyperref}
\usepackage{booktabs}
\usepackage{subfigure}
\usepackage{xcolor}

\definecolor{Blue}{rgb}{0,0,1}

\begin{document}

\title{Interconnecting Regional QKD Networks: Hybrid Key Delivery Across
  Quantum Domains\thanks{This work was supported by MICINN through the European Union
  NextGenerationEU recovery plan (PRTR-C17.I1), and by the Galician Regional
  Government through the \textit{Planes Complementarios de I+D+I con las
    Comunidades Autónomas} in Quantum Communication, and by the Basque
  Government through \textit{Plan Complementario Comunicación Cuántica}
  (EXP. 2022/01341) (A/20220551). The authors also acknowledge Galicia
  Supercomputing Center (CESGA) for providing access to the FinisTerrae III
  supercomputer with funding from the \textit{Programa Operativo
    Plurirregional de España} 2014-2020 of ERDF, ICTS-2019-02-CESGA-3. Iago
  Fernández Llovo acknowledges support from Xunta de Galicia through grant
  IN606B-2025/027.}
  }

\author{David Barral, Aitor Brazaola-Vicario, Diego Cifrián, Natalia Costas,
  Gonzalo Blázquez, Ana Fernández-Vilas, Iago F. Llovo, Pedro Otero-García,
  Pablo P. Rejo, Alejandra Ruiz, Juan Villasuso, Manuel Fernández-Veiga}

\maketitle

\begin{abstract}
   QKD technology is being increasingly adopted inside the network core for 
   protecting information transport against any form of computational attacks. However,
   the use of QKD for wide-area internetworking is still challenging and costly, due to 
   its strong trust assumptions and the low achievable key rates in long QKD links. This 
   paper presents a standards-driven design and implementation of a unified hybrid 
   key delivery service for a network of isolated QKD domains (subnetworks using 
   QKD as provider technology for secret key generation) connected via classical WAN
   links. The framework follows a distributed architecture and uses a hybrid approach where 
   keys generated in a domain are securely relayed to other domains with PQC (Kyber), routed 
   through automatic topology-aware path computation, and managed at the system level. The solution 
   has been implemented in an operational testbed comprising three regional subnetworks. We present
   the design principles, the deployment, and the experimental performance results for
   this scalable key delivery service.
\end{abstract}

\section{Introduction}
\label{sec:introduction}

Quantum Key Distribution (QKD) has rapidly become a practical and mature technology to 
achieve information-theoretical security both in isolated point-to-point links spanning up 
to tens of kilometres and in long-haul multi-hop networks, using trusted nodes that interconnect
adjacent QKD links for relaying the quantum-generated keys across a network 
path~\cite{Stanley2022}. This architecture for distributing the keys between two endpoints
replicates and leverages the physical and logical infrastructure of the optical backbone in 
the current Internet. So, the control plane that provides the key management service  and the data 
plane carrying the encrypted information substantially overlap and, 
in many cases, share communication resources~\cite{Aguado2019}. The integration 
of QKD technology  into the carrier infrastructure of a network operator is recognized as 
an essential evolutionary path toward the deployment  of QKD-secured communications on 
wide-area networks~\cite{Avesani22}.

However, tightly coupling the key management and the network data plane is not necessarily
the best architectural approach to develop an efficient, stable, and general key 
distribution service for a large-scale QKD network, for several reasons~\cite{Tsai2021}. 
First, QKD technology is still expensive, specialized, and complex to operate, with 
limited points of presence in telecommunications networks~\cite{Mehic2020}. Therefore, 
it is unrealistic to foresee a configuration where a majority of the end nodes or 
end-user devices have direct access to QKD nodes closely located to their premises, 
to obtain from them the raw keys required by the application layer~\cite{Aquina2025}. 
Similarly, deploying QKD equipment in a significant fraction of links in the backhaul 
network will be a complex and costly task, and the proliferation of trusted relay nodes 
exposes the vulnerability of the  Key Management System (KMS)~\cite{Cao2022}. A second problem 
is that the  performance of QKD in terms of secret key rate (SKR)  is highly variable, since 
the physical process is very sensitive to the specific quantum protocol used to generate 
the keys~\cite{Scarani2009}, to the link distance~\cite{Pirandola2017}, and to the
imperfections in the devices and the links. As a result, a large QKD network has to 
face challenges related to the variability and heterogeneity in the key generation
process~\cite{Cao2022}. Specifically, QKD devices do not generate the keys at constant,
predictable rates, but rather in discrete bursts at irregular intervals, after enough 
secret bits have been distilled from the quantum measurements and have been packed 
as a block~\cite{Islam2017}. This time-varying behaviour further complicates the 
allocation of keys  to nodes or devices, and makes it difficult to estimate the SKR 
of the networks as a  whole. Besides, a common assumption is that key material on 
QKD nodes is immediately available to any arbitrary user. But such on-demand 
consumption can easily deplete the stored key material, and lead to  dramatic performance 
drop or unfair usage of  the keys among distinct applications~\cite{Dervisevic2025}.

Accordingly, it seems likely that, in the near future, QKD networks will be gradually 
deployed and operated on smaller areas, realizing a sparse collection of autonomous
QKD sub-networks or QKD islands, bridged with state-of-the-art non-QKD
WAN technology~\cite{Cao2022,Li2025}. Hence, a KMS designed for a QKD network implemented 
on almost identical devices, which also relies on the principles of constant key supply 
and demand, is clearly inadequate to satisfy the requirements and key management performance 
of a decentralized collection of KMSs interconnected through non-QKD 
links~\cite{Dervisevic2025}. A more flexible architecture is required, that guarantees (i) 
confidentiality in the key-relaying process; (ii) interoperability among the individual KMSs 
used in each subnetwork; (iii)  reliable/predictable performance for the key delivery to the application 
layer, in a hardware-independent and topology-independent form; (iv) openness and 
interoperability across different administrative network domains or autonomous systems.

This paper presents the design, implementation, and performance of a distributed KMS
specifically designed to address the aforementioned challenges. Our main contributions in 
this  work are the following:
\begin{enumerate}
\item We design and implement a distributed KMS for an arbitrary federation of QKD sub-networks, 
which can be connected with QKD or non-QKD links. The proposed architecture extends ETSI-compliant 
trusted-node operation to the interconnection of geographically separated QKD domains through hybrid QKD--PQC 
key delivery. Our implementation is open-source and is compliant to the standard API interfaces ETSI GS QKD 
014~\cite{ETSI014} \& ETSI GS QKD 020~\cite{ETSI020}. The southbound interface in the former reads out the 
key from the internal KMSs of the QKD devices, while the latter is used to coordinate the exchange of keys 
among the nodes that form the overlay KMS network.\footnote{The source code for our implementation can be 
found in Zenodo, with DOI \url{10.5281/zenodo.20713365}.}

\item To solve the secure propagation of keys over non-QKD links, we adopt a hybrid 
key transport scheme that encrypts the key material with Post-Quantum Cryptography (PQC). 
In particular, we use Kyber (FIPS 203~\cite{NIST2024}) for post-quantum protection in 
transport and Transport-Layer Security (TLS) from a KMS sender to a KMS receiver. Since a 
pure QKD path with ITS does not exist between the subnetworks, and models like simple 
key-relaying (KR) and trusted-node (TN)~\cite{ITUT2020} do not ensure end-to-end 
confidentiality across intermediate nodes, this hybrid QKD plus PQC approach is required 
for comprehensive security of the KMS.

\item We deploy our complete KMS on a real testbed connecting four sites running
in two Spanish regional QKD networks, separated by 500 km. One of the sub-network includes 
three sites with QKD links, one of which is the longest QKD link in operation in Spain 
($\sim 120$ km). Our distributed KMS demonstrates topology-aware routing of key requests 
through automatic path computation across the distributed KMS infrastructure. Also, we conducted 
an extensive set of performance measurements on this testbed (delay, SKR, number of keys exchanged, 
correlations) and report our experimental findings and best practices on this system. 
\end{enumerate}

The rest of this paper is structured as follows. Some background and a review of the works
related to this paper is given in Section~\ref{sec:background}. The system architecture,
design principles and implementation decisions of the KMS are presented in 
Section~\ref{sec:architecture}. Next, Section~\ref{sec:key-relaying} explains in detail the 
procedure devised for secretly forwarding the keys hop-by-hop through the key management nodes.
The experimental setup of our network is described in Section~\ref{sec:testbed}. 
Section~\ref{sec:results} reports the experimental results obtained with the testbed, and 
these are discussed in Section~\ref{sec:discussion}. Finally, Section~\ref{sec:conclusion} 
contains the conclusions and summary of the paper.

\section{Background and related work}
\label{sec:background}

Large-scale QKD pilots and national initiatives have progressively moved QKD from laboratory 
demonstrations to operational networked services, where the central challenge is no longer limited 
to key generation but extends to the scalable handling, interoperability, and delivery of key material 
across heterogeneous infrastructures. As deployments expand geographically, architectural integration 
and flexible key management become decisive factors for sustainable growth.

\subsection{From pilot QKD networks to federated infrastructures}

In the early 2000s, the first networked QKD experiments demonstrated that quantum-secured 
communication could operate reliably over deployed fibre infrastructures while revealing the
architectural requirements for extending reach and integrating applications. In 2003, the DARPA 
Quantum Network~\cite{Elliott2003} provided one of the earliest sustained metropolitan-scale 
deployments, establishing practical operational concepts that influenced subsequent 
architectures. A few years later, in 2007, the SECOQC network in Vienna~\cite{Dianati2007} 
advanced a multi-vendor trusted-repeater model and formalised the separation between quantum 
devices, Key Management Systems, and applications. 

By 2011, both the Tokyo QKD Network~\cite{Fujiwara2011} and 
SwissQuantum~\cite{Stucki2011SwissQuantum} demonstrated complementary aspects of networked 
QKD maturity. The Tokyo deployment showcased heterogeneous QKD technologies within a mesh
topology, highlighting dynamic reconfiguration and application-driven operation, while 
SwissQuantum focused on long-term stability and carrier-grade robustness.

As the following decade progressed, attention shifted towards national and continental 
infrastructures. The Beijing--Shanghai backbone, whose operational phase was reported in 
2021~\cite{Chen2021BeijingShanghai}, operationalised trusted-node chaining over thousands of 
kilometres by combining terrestrial fibre and satellite segments. In Europe, OPENQKD (2018–
2022)~\cite{OpenQKDNetwork} and the EuroQCI initiative, launched in 2019~\cite{euroQCI}, have 
pursued the federation of regional infrastructures into a continental quantum communication 
ecosystem. More recently, in 2023 the Madrid QKD testbed~\cite{Martin2023} and in 2024 
cross-border European interconnection efforts~\cite{Brauer2024} illustrated the transition
towards software-defined and orchestrated quantum networking~\cite{Aguado2019,Lopez2025}. 
Architectural analyses indicate that, as networks mature, challenges increasingly shift 
from quantum transmission performance itself to orchestration and inter-domain 
coordination~\cite{Mehic2020,Cao2022,Stanley2022}.

Across these initiatives, intermediate TNs retrieve, synchronise, and forward key material 
hop by hop. While this model enables scalability beyond direct fibre reach, it often presumes
homogeneous QKD connectivity between adjacent nodes, which can constrain expansion when 
linking geographically separated QKD “islands”.

\subsection{Trusted nodes, KMS interoperability, and ETSI interfaces}

The scalability of TN networks depends critically on the key management layer. KMS instances 
coordinate synchronisation, lifecycle management, and the exchange of key material between 
heterogeneous QKD devices. Foundational analyses of TN security~\cite{Salvail2010} and more 
recent surveys on QKD network architectures and key 
management~\cite{Lella2023,Dervisevic2025,Berl25} converge on the importance of robust and 
interoperable KMS designs.

To support interoperability, ETSI has defined a family of QKD key delivery specifications. 
ETSI~GS~QKD~004~\cite{ETSI004} specifies an application-facing key supply interface, while 
ETSI~GS~QKD~014~\cite{ETSI014} defines a REST-based API used for key requests, retrieval, and 
status exchange between KMSs and Secure Application Entities. ETSI~GS~QKD~020~\cite{ETSI020} 
provides a standardised KMS-to-KMS interface that structures how key-related information is 
coordinated between KMS instances. Together, these interfaces define a common control and 
data plane vocabulary for interoperable deployments, as summarised in 
Table~\ref{tab:etsi-interfaces}. They underpin infrastructures such as 
OpenQKD~\cite{OpenQKDNetwork} and KMS prototypes including 
QCI-CAT~\cite{qcicatManagementSystem}.

\begin{table*}[t]
  \centering
  \caption{Relevant ETSI QKD interfaces referenced in this work}
  \label{tab:etsi-interfaces}
  \begin{tabular}{@{}lll@{}}
    \toprule
    \textbf{Specification} & \textbf{Scope} & \textbf{Description} \\ \midrule
    ETSI~GS~QKD~004 & Key supply & Implementation-agnostic, session oriented. \\
    ETSI~GS~QKD~014 & Key supply & REST-based key delivery between KMS and SAE. \\
    ETSI~GS~QKD~020 & KMS coordination & Interoperable KMS-to-KMS API for coordinating key 
    material. \\
    \bottomrule 
  \end{tabular}
\end{table*}

\subsection{Hybrid QKD--PQC approaches and positioning of this work}

As QKD networks evolve into federations of sub-networks, hybrid approaches are increasingly 
explored to ensure quantum-safe operation across segments where QKD links are unavailable or
impractical.

In 2021, the Jinan metropolitan deployment~\cite{Yang2021} replaced the traditional symmetric 
pre-shared authentication of the classical QKD channel with a lattice-based post-quantum 
signature scheme integrated into a commercial QKD system, demonstrating stable operation in 
a 14-node metropolitan field network. In 2023, Geitz \emph{et al.}~\cite{Geitz2023} 
implemented a quantum-secure Key Management  System within the OpenQKD Berlin testbed, 
replacing classical  Rivest--Shamir--Adleman (RSA) \cite{Rivest1978} and Elliptic Curve 
Cryptography (ECC)~\cite{Koblitz1987} mechanisms with PQC key encapsulation and signature 
schemes. Their approach hardened node authentication and key forwarding between QKD nodes 
but relied on a dedicated PQC-based KMS design rather than ETSI-compliant interoperability.

Doering \emph{et al.}~\cite{Doering2022} proposed integrating a PQC key exchange mechanism 
that mimics QKD behaviour within an existing QKD platform, allowing both QKD and PQC keys to
coexist in the same secure key store and enabling applications to distinguish between them 
via metadata. At the protocol level, García \emph{et al.}~\cite{Garcia2024} investigated 
hybrid QKD–PQC authenticated key exchange mechanisms embedded within TLS, evaluating 
performance and security trade-offs of concatenation and XOR-based designs. In parallel, 
Garms \emph{et al.}~\cite{Garms2024} introduced a hybrid authenticated key exchange 
cryptosystem combining QKD, PQC and physical unclonable functions (PUF), implemented on 
Field-Programmable Gate Array (FPGA) hardware as an integrated quantum-safe architecture.
More recent system-level discussions emphasise that quantum-safe networking will likely 
rely on structured combinations of QKD and PQC rather than on either technique 
alone~\cite{Mehic2024,RubioGarcia2025}.

Table~\ref{tab:relatedwork-comparison} compares these representative efforts with the architectural 
focus of this work. While existing studies demonstrate hybridisation at the level of authentication, 
protocol integration, or cryptosystem design, fewer contributions address the trusted-node architecture 
itself as the integration point for heterogeneous secure links.

\begin{table*}[t]
\centering
\caption{Comparison of closely related QKD--PQC efforts and the focus of this work}
\label{tab:relatedwork-comparison}
\footnotesize
\begin{tabular}{p{1.3cm}p{7.6cm}p{2.7cm}p{1cm}p{1cm}}
\toprule
\textbf{Work} & \textbf{Primary focus Evidence} & \textbf{Layer} & \textbf{KMS APIs} & \textbf{QKD islands} \\
\midrule

\cite{Yang2021} &
PQC-based authentication of QKD classical channel replacing symmetric pre-shared authentication;
Metropolitan field deployment &
\raggedright Control plane authentication &
\texttimes &
\texttimes \\

\cite{Geitz2023} &
Quantum-secure KMS using PQC key encapsulation and signatures for node authentication and key forwarding; 
OpenQKD testbed implementation &
\raggedright Key management system implementation &
\texttimes 
&
\texttimes \\

\cite{Doering2022} &
Integration of PQC key exchange mimicking QKD behaviour within a QKD platform; Prototype integration in OpenQKD &
\raggedright Access and transport integration &
\texttimes &
\texttimes \\

\cite{Garcia2024} &
Hybrid QKD–PQC authenticated key exchange within TLS; Experimental TLS implementation &
\raggedright Transport security protocol layer &
\texttimes &
\texttimes \\

\cite{Garms2024} &
Hybrid authenticated key exchange cryptosystem combining QKD, PQC and PUFs; FPGA-based hardware prototype &
\raggedright Cryptosystem architecture and hardware integration &
\texttimes &
\texttimes \\

This work &
Flexible trusted-node architecture integrating QKD links and PQC-protected segments under ETSI-compliant 
KMS interfaces; Regional multi-site deployment &
\raggedright Trusted node and key management architecture &
\checkmark 
&
\checkmark \\
\bottomrule
\end{tabular}
\end{table*}

Against this background, our contribution is framed at the level of the Trusted Node architecture 
itself. Rather than introducing isolated protection mechanisms, we design a KMS-centric architecture 
that builds upon ETSI-compliant interfaces and exploits the interoperable message structures summarised 
in Table~\ref{tab:etsi-interfaces}. The resulting framework extends ETSI GS QKD 020 beyond conventional 
trusted-node operation to enable secure key delivery across geographically separated QKD domains 
interconnected through classical WAN segments. By combining QKD-protected and PQC-protected links within 
a unified standards-compliant framework, it provides interoperability between heterogeneous QKD islands 
and is validated through a multi-regional deployment spanning approximately 500 km. The architecture and 
its integration mechanisms are described in the following section.

\section{System architecture} 
\label{sec:architecture}

\subsection{Service Elements}
To achieve scalable quantum-resistant key delivery in next-generation networks, we propose a Key 
Distribution Service constructed from a set of distributed Key Management Systems—defined 
in the ETSI-QKD framework as Key Management Entities (KMEs). These distributed entities 
collaborate within a unified key management layer to orchestrate the lifecycle and delivery 
of cryptographic material across complex topologies. Recognizing the physical constraints 
of quantum signals, this architecture is designed to bridge distinct network segments by 
combining point-to-point Quantum Key Distribution links with classical communication 
segments. To ensure long-term resilience against quantum threats across the entire 
infrastructure, the service adopts a hybrid approach: it leverages QKD for 
information-theoretic key establishment where physical links are available, while integrating 
Post-Quantum Cryptography to secure segments where QKD links are not available. This 
synergy ensures that the distributed KMS can facilitate multi-hop key relaying via trusted 
nodes, maintaining a continuous chain of trust. This Section describes the system architecture, 
its main components and the operational workflow of the key generation and distribution process.  

The design aims at a versatile key distribution service (KDS) system comprised of: 
\begin{enumerate}
\item SAEs (Secure Application Entities)  represent the end-points of the cryptographic 
ecosystem, that consume cryptographic material to establish secure communications. Within 
the proposed architecture, SAEs function as the clients of the  KDS. Instead of generating 
their own keys, SAEs utilize standardized interfaces to request and retrieve synchronized 
key material from their local KMS. In a typical workflow, an initiating SAE (Master) requests
keys for a specific session with a peer SAE (Slave). The SAEs rely on the underlying KMS to
transparently handle the complexity of key generation, synchronization, and multi-hop 
relaying across the network. By decoupling the key usage from key generation, the SAEs
ensure that the KDS delivers keys that are legally distinct from the payload data, adhering
to the principle of separation of mechanism and policy.

\item KMSTNs (Key Management System Trusted Node) function as the Key Management Entities
defined in ETSI  GS QKD 004 and 014, i.e. KMSTN acts as the software agent that interfaces
between the end-user  SAEs and the physical QKD infrastructure. Operating within the trusted 
node boundary, the KMSTN is responsible for the lifecycle management, synchronization, and
delivery of quantum-generated keys. Beyond managing local links, these entities collaborate 
to form a distributed key management layer capable of sharing keys beyond the distance limit 
of a single QKD link. To this end, KMSTNs execute two major tasks: first, they cooperate to
discover network paths for key relaying between source and destination SAEs; second, they
securely forward keys across the network-encapsulated by PQC or a hybrid QKD+PQC scheme
---when adjacent trusted nodes communicate via classical links, ensuring a continuous 
chain of trust.

To overcome the point-to-point physical limitations of quantum signals, KMSTNs do not 
operate in isolation; they communicate with peer KMSTNs to form a distributed Key 
Management Layer. This inter-node communication is essential for two functions:
\begin{enumerate} 
\item Key Synchronization and Association: For adjacent nodes connected by a physical 
quantum channel, the KMSTNs interact to post-process and synchronize the raw key material,
establishing a Direct Key Association Link. 

\item Key Relay and Virtual Links: To serve distant SAEs, KMSTNs cooperate to establish 
Virtual Links. Through a hop-by-hop Key Relay process, a KMSTN receives a key from a 
previous node (encrypted with a local key) and forwards it to the next node in the path. 
This mechanism allows the dynamic provisioning of end-to-end keys across the network 
topology without requiring a direct optical path between the endpoints. 
\end{enumerate}

\item Network controller: It is the 
controller of the software-defined cryptography infrastructure (SD-QKD), providing a global 
view of both the quantum and classical layers. Its operation is governed by SDN control 
interface standards (e.g., ETSI GS QKD 015) and focuses on three main functions: 
(1) Topology Discovery and Resource Management to maintain a real-time abstraction of the 
network topology; (2)  Path Computation and Service Provisioning for end-to-end connectivity
between two distant SAEs (configuration of intermediate trusted nodes to establish virtual
links); and Application Registration and QoS Enforcement. As described later, these logical
functions of a network controller are embedded in our implementation into the KMSTN for
practical issues, yet they stand generally as an independent functional and operational 
module.
\end{enumerate}

\subsection{Security Model}

Our first assumption to analyze the security of the key distribution service is
that the generation of the QKD keys in a quantum link is information-theoretically 
secure, and that the QKD nodes cannot be controlled or hacked by an adversary. Upon
this premise, the remaining elements in the architecture make use of the following 
assumptions and mechanisms to fulfil security against attackers able to intercept, 
modify, and change any message on the classical channels.
 
\begin{enumerate}
\item \textbf{Authenticated and encrypted vertical channels}. As per the ETSI QKD GS 
014, we use mutual TLS in the links used by those protocols. Specifically, it is only 
on its own for the ETSI GS QKD 014 API communication between KMSTN and the QKD machines 
and between KMSTN and the SAEs. 

\item \textbf{Authenticated and encrypted horizontal channels}. ETSI GS QKD 020
messages are exchanged over classical links and may transport key material.
Therefore, protecting their confidentiality and integrity is essential. Although
the specification mandates HTTPS with mutual TLS authentication, key relaying is
the fundamental service provided by the KDS and we therefore protect ETSI GS QKD
020 messages at the application layer through an additional encryption mechanism,
independent of the underlying TLS session. This protection is implemented using AES-256-GCM, 
with session keys derived from fresh Kyber shared secrets or, when available, from a hybrid 
combination of Kyber and QKD key material. The session-key generation process is summarised
in the following equation
\begin{equation}
\label{eq:aes-key-derivation}
K_{\mathrm{AES}} =
K_{\mathrm{Kyber}}
\oplus
K_{\mathrm{QKD}}^{*}.
\end{equation}
Here, $K_{\mathrm{QKD}}^{*}=0$ for PQC-only links and
$K_{\mathrm{QKD}}^{*}=K_{\mathrm{QKD}}$ when a QKD link is available,
enabling the hybrid QKD--PQC protection described above.

A fresh Kyber exchange is performed for each protected message and a fresh
96-bit initialization vector (IV) is generated for every AES-256-GCM encryption operation. By enabling
secure transport across WAN segments where direct QKD connectivity is
unavailable, the hybrid approach allows geographically separated QKD domains to
operate as a unified key-distribution infrastructure.

The resulting security guarantee is therefore path-dependent. When every horizontal segment involved in the delivery path is protected using QKD-derived secret material, the transport can retain an information-theoretic component, subject to the trusted-node assumptions. Conversely, if the path includes at least one PQC-only segment, the confidentiality of the delivered key material is computational rather than information-theoretic and is bounded by the security of ML-KEM/Kyber and AES-256-GCM on that segment. Accordingly, the proposed architecture must not be interpreted as providing unconditional end-to-end security across heterogeneous QKD–PQC paths.

\item \textbf{Key relaying}. This is generally implemented through OTP (One Time Pad)
between the received key and a locally generated (and secure) key, used only once. OTP
provides perfect encryption security (in an information-theoretic interpretation).
We have used a variation of OTP that hybridizes QKD and PQC by doing the XOR between 
QKD delivered keys and Kyber delivered secrets. Such compound keys can only be recovered 
by an attacker if both keys are eventually compromised. This approach mitigates the risks
due to potential security holes in the classical software and hardware used inside the
QKD nodes to post-process and deliver the consumable keys. It also helps to reduce
possible vulnerabilities to Kyber (or other similar lattice-based PQC encryption 
algorithms) by combining the Kyber KEM output with a genuine random QKD key. 

\item \textbf{Hardware-backed secure key storage}. During runtime, KMSTNs retrieve
keys from the underlying QKD nodes and store them in local databases. To protect
key material at rest, we leverage the Trusted Platform Module (TPM) available in
most modern computers and a SQLCipher database encryped using AES-256-CBC. 
The TPM is a secure cryptoprocessor that we use for  encryption and decryption 
with a key that never leaves the TPM and cannot be directly extracted by the host 
operating system. However, a TPM is not a Trusted Execution Environment (TEE) 
and therefore does not protect application memory or runtime execution.

In the implementation (see Section~\ref{sec:testbed}), the strongest AES variant
supported by the TPM is automatically selected. 
The SQLCipher master password is stored on disk wrapped using AES-CFB under 
a non-exportable TPM-resident key and decrypted by the TPM at runtime, so the 
credential is never persisted in plaintext or exposed in environment variables, and 
the encrypted secret is bound to the host. The field of the database that stores the 
QKD Keys  is encrypted through the TPM using AES-CFB and  it is decrypted by the TPM 
on access rather than when the database is loaded; the corresponding encryption key 
never enters application memory, and QKD Key plaintext is materialized just-in-time only 
for accessed fields. The master password is briefly present in memory while the 
database is instantiated; The protection of the database password is therefore against 
at-rest and offline credential disclosure rather than live memory compromise and it 
does not provide the runtime isolation guarantees typically associated with TEEs.
\end{enumerate}

\subsection{Adversarial Model, Trust Assumptions and Threats Mitigated}

Once the main security mechanisms have been presented, we identify the threats against which 
the system defends, characterize the conditions under which these defenses hold, and acknowledge the 
residual risks that remain outside the scope of cryptographic enforcement. Specifically, we consider a 
standard computational adversary $\mathcal{A}$,  modelled as a PPT algorithm consistent to the Dolev-Yao 
threat model~\cite{Dolev1983}, extended for quantum-safe  operations. The capabilities of $\mathcal{A}$ include:
\begin{itemize}
\item \textbf{Network control}. $\mathcal{A}$ observes and may delay, drop, or
  reorder all messages exchanged between entities (SAEs and
  KMTNs). $\mathcal{A}$ cannot break the underlying transport-layer
  cryptography assumed for the communication channels between a SAE and a
  KMSTN, or between two KMSTNs, except with negligible probability in the
  security parameter $\lambda$.

\item \textbf{CRQC control} Under CRQC (Cryptographically Relevant Quantum
  Computing), $\mathcal{A}$ is capable of running Shor’s algorithm to break
  classical public-key cryptography (e.g., RSA, ECC) in polynomial time, and
  Grover’s algorithm to accelerate symmetric key searches.

\item \textbf{Storage \& user control}. $\mathcal{A}$ cannot extract keys out
  of the TPM, but can compromise a subset $\mathcal{C}$ of the KMSTNs and
  read/write the memory.
\end{itemize}
We can now analyse the principal threats that the system defends against, in
each case naming the structural mechanism on which the defense rests.
\begin{enumerate}
\item 128-bit Quantum-Resilient Confidentiality is achieved with the use of
  256-bit AEAD symmetric ciphers (AES-256-GCM and AES-GCM-CBC) at the TLS 1.3
  data plane. This maintains a secure 128-bit classical equivalent against
  quantum search algorithms like Grover’s, ensuring the long-term secrecy of
  user and control plane payloads.  The authenticated encryption is secure
  against chosen-ciphertext attacks. Quantum-resilient confidentiality at this
  strength is ensured both for SAE to KMSTN communications and for inter-KMSTN
  communications as well.

\item Perfect Forward Secrecy is achieved on classical channels with the use
  of an ephemeral post-quantum KEM (Kyber) in a hybrid configuration with
  classical encryption (AES-256-GCM) for downstream key relaying between
  KMSTNs. The AES key generated via~\eqref{eq:aes-key-derivation} is used only
  once for each message between the KMSTNs. On the KMSTNs, forward secrecy is
  ensured against any leakage of the database contents, since the keys are
  hardware-encrypted with the TPM and can only be decrypted in the same
  device.

\item Classical-Resilient Authentication is guaranteed by the use of mutual
  TLS authentication between peer entities. This mechanism prevents
  impersonation or forgery attacks, i.e., it is secure against EUF-CMA.
  Quantum-resilient authentication can be enforced by replacing X.509
  certificates with PQC-signed certificates, e.g. with ML-DSA.

\item Information-theoretic security of locally generated QKD key
  material. The key material produced over a physical QKD link inherits the
  information-theoretic security guarantees of the underlying QKD system,
  under the assumptions stated above. This guarantee applies to key generation
  and to transport over paths composed exclusively of appropriately secured
  QKD links. It does not extend end to end to a relaying path containing a
  PQC-only horizontal segment. In such paths, confidentiality is computational
  and depends on the security of ML-KEM/Kyber and AES-256-GCM.
\end{enumerate}

No cryptographic protocol can defend against threats that lie outside the
scope of the abstractions it formalizes.  The main residual risk uncovered by
our design is the compromise of a set $\mathcal{C}$ of KMSTNs. This would
allow $\mathcal{A}$ to inspect the keys in the database and the forwarded
keys. The cryptographic layer offers no defense against this, because the
KMSTN is crucial in the trust model. The practical mitigation of such risk is
to strengthen the runtime execution environment of the KMSTNs and upgrade it
into a fully TEE.

\section{Key relaying mechanism \& protocol} 
\label{sec:key-relaying}

KMSTNs can be deployed in an unconstrained topology overlaid to the physical QKD
infrastructure, provided that a subset of the KMSTN nodes are connected to existing
QKD nodes (to their KMS entities, to be precise) in order to request and consume the
quantum keys generated at the physical layer. The southbound API between a KMSTN and 
a SAE or a QKD KMS is the standardized ETSI GS QKD 014, allowing the system to be
transparent and agnostic of the QKD equipment, vendor or protocol in use in the lower 
quantum layer. Using the same API to receive requests from SAEs underlines the role
of KMSTNs as providers of end-to-end keys for the applications, acting as secure 
proxies between these and the real QKD infrastructure. Thus, SAEs do not need to have 
direct access to QKD nodes but can invoke the use of quantum keys to secure the 
communications with a peer SAE.

The coordination between KMSTNs follows the ETSI GS QKD 020 specification for key 
relaying. As of today, this specification is still a draft protocol that works out
the horizontal exchange of QKD keys between two systems located in a common trusted 
node. It follows a proactive key sharing model in which the keys are relayed from one 
node to the next hop. It also enables the emulation of QKD protocols and key exchanges 
even when no underlying QKD infrastructure exists, so it fulfils the functional 
requirements for key distribution over a sparse and discontinuous QKD network topology. 

\begin{figure}[!t]
  \centering
  \includegraphics[width=0.95\textwidth]{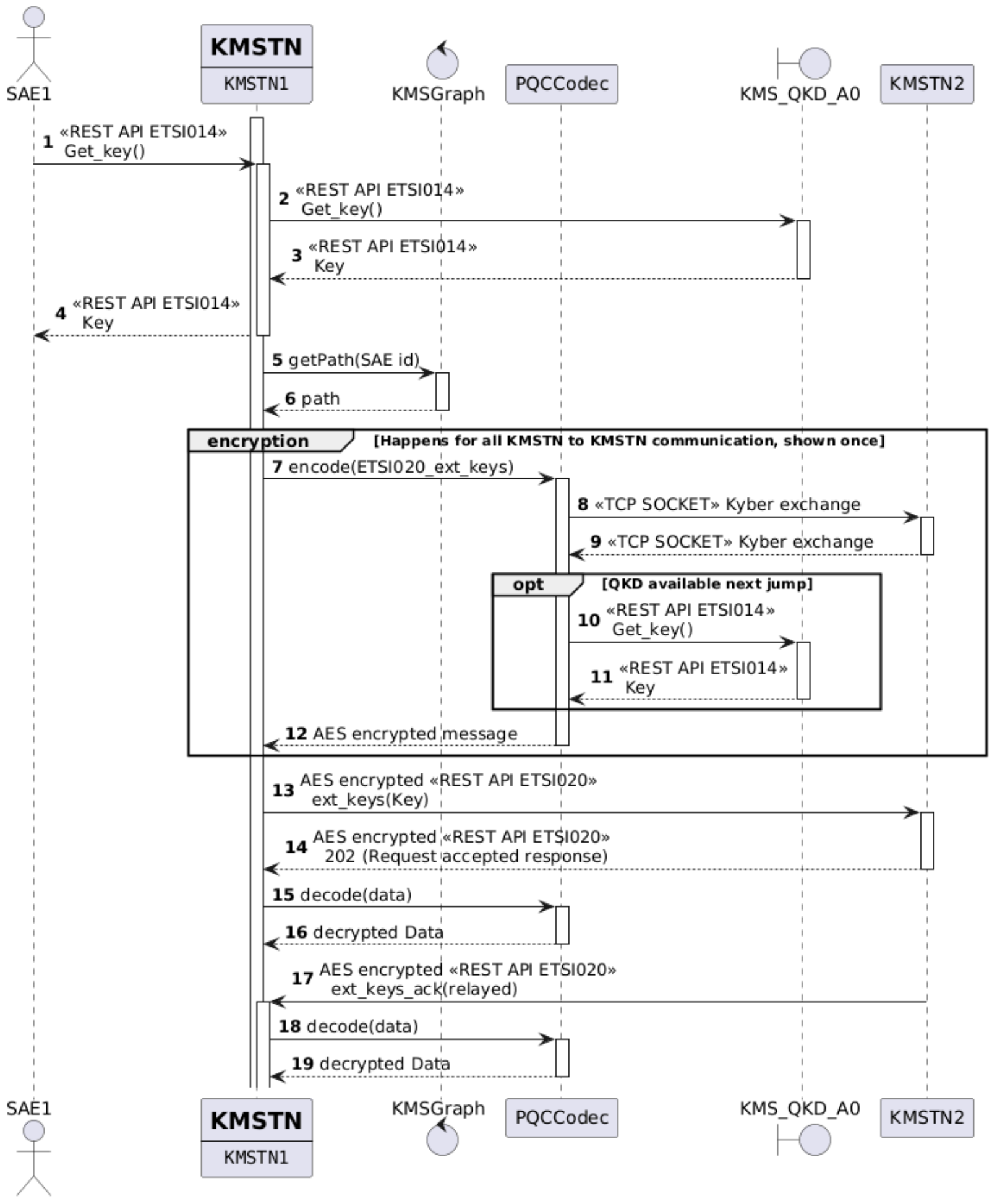}
  \caption{\label{fig:get-key-with-ids} Workflow and message exchange between
    the KMSTN entities and the SAEs for a key request.}
\end{figure}

The main function of this software is routing keys across multiple KMS to extend the 
reach of QKD, provide multicast when the QKD provider does not and possibly provide
redundancy. To achieve that, we proactively export keys using ETSI GS QKD 020 and route 
them from one KMS to another until they reach their intended receiver using QKD + PQC or 
PQC alone when no QKD is available to secure the routed keys. In the following, we will 
detail the process step by step. An example of the protocol coordination and message exchange
for an end-to-end key request operation and forwarding is depicted in 
Figure~\ref{fig:get-key-with-ids}.

\subsection{End-to-end key negotiation process}

\subsubsection{Path computation}The initiating Master SAE requests keys from the KMSTN by
invoking the ETSI GS QKD 014 REST API. While the request must target a local QKD node 
(Alice), the Master SAE identifies the ultimate destination by including the \textsc{sae\_id}
of the remote peer (Slave SAE). Upon receiving this request, the KMSTN identifies the 
appropriate local QKD interface and issues a corresponding key request via the southbound 
API. Crucially, the KMSTN acts as a proxy that sanitizes this forwarded request: it 
replicates the original parameters but omits the additional slave \textsc{sae\_id}s, 
as the native QKD node lacks awareness of the overlay topology and would reject a request 
containing unknown remote identifiers. Once the native QKD node generates and returns 
the key, the KMSTN delivers it to the Master SAE.  Concurrently, the KMSTN utilizes 
the previously protected slave \textsc{sae\_id} to identify the next hop within the network graph 
of the KMSs, and securely forwards the key material to the corresponding peer KMSTN. 

\subsubsection{Key forwarding}

The procedure to send the key from the origin KMSTN to the destination KMSTN goes 
as follows. Keys are encapsulated within \textsc{post\_ex} messages specified in the 
ETSI GS QKD 020 standard. The sending KMSTN first checks if there exists a QKD link that
can be used to secure the route between the current KMSTN and the next. If so, it requests 
a QKD key and next, upon success, sets the Kyber ID equal to the QKD key ID to start 
the Kyber encapsulation  protocol and agree a secret value. When the Kyber KEM is
completed, the ETSI GS QKD 020 \textsc{post\_ex} message is encrypted with AES-256-GCM, along with
the ID and the IV.

The receiving KMSTN in this hop, in response to the incoming message, checks first whether
a QKD link exists and has been used between the upstream KMSTN and itself. In such a case,
the local QKD node (Bob) is asked about the key ID, which has been previously decrypted
from the received message. The KMSTN node obtains the key value $k_{AB}$ and immediately
performs the OTP $k^\prime_{AB} := k_{AB} \oplus \operatorname{KEM}_{AB}$ with the
Kyber-generated secret value. The OTP $k^\prime_{AB}$ is now the message key to be f
forwarded to the next-hop KMSTN. 

If a QKD link does not exist ---so the incoming message has been transmitted encrypted 
with PQC over a classical link--- the receiving KMSTN sets the message key to the
value of the Kyber secret agreed with its upstream partner, decrypts the message with
AES-256-GCM (using the message key and the IV included in the message), and checks if any 
of the the \textsc{sae\_id}s specified in the Key matches its own master \textsc{sae\_id}. 
The key is then  stored in its KeyStore database if the test succeeds, otherwise it 
verifies whether a match happens with any other KMSTN master \textsc{sae\_id} and, 
if this holds,  the routing procedure is repeated.

When, at a later moment, a SAE requests a key to a KMSTN, it must specify the
key ID obtained by its peer SAE at the beginning and also the slave \textsc{sae\_id} of 
the QKD node Alice in which this key was generated. How the key ID is communicated 
between SAEs is beyond the scope or control of our protocol. The KMSTN uses queries
its KMSGraph to retrieve the KMSs that bind the specified \textsc{sae\_id} as their 
master SAE;  now, supposing that the master SAE is one for which this KMSTN has a direct 
connection to, the key is requested directly through a ETSI GS QKD 014 call.  If such 
direct connection does no exist, then the KMSTN attempts to retrieve it from its key 
store. The key is then returned to the requesting SAE, or an error is finally returned 
if the key cannot be found. 

\subsection{Secret exchange using Kyber} 

Each KMSTN exposes a socket listening for Kyber secret exchanges. The sequence 
starts with one client sending a public key request to the raw socket in the server. 
The payload is the binary representation of the string \texttt{REQUEST\_PUBLIC\_KEY}. 
The server will reply with a message that identifies it as a public key reply. The 
payload contains the public key of the server. Then the client will use the server's 
public key to produce a ciphertext and will send it to the server's socket. It has a
payload with the binary representation of the ciphertext.  The server receives the 
message with the ciphertext and uses its private key to recover the secret that was 
used by the client to generate it. It will store it associated with the 
\texttt{session\_id} which can be communicated by any other protocol that needs to use 
the secret they exchanged.   
 
\subsubsection{Session initialization and TLS handshake} 

We use standard mutual TLS (mTLS) as ETSI GS QKD 014 and ETSI GS QKD 020 stipulate. 
However, that is a minimum; standard TLS is not considered PQC and does reuse keys 
during sessions, making it potentially vulnerable to statistics-based attacks. We rely on 
mTLS to secure the 014 API messages, as they are meant to happen inside of secured 
networks between the SAE and the QKD or KMSTN. The same is true for KMSTN to QKD 
communication. However, on KMSTN to KMSTN communication using 020 while we also use mTLS, 
we do not rely on it as the  messages have their own encryption layer on top of mTLS in which 
the keys are purely ephemeral and never reused. 

\subsubsection{Secure Key Storage and retrieval}

As ETSI GS QKD 020 is a proactive key sharing protocol, keys may need to be stored for 
any arbitrary lengths of time until a SAE requests them by their ID. To that end, 
we have used a database (SQLcypher) that includes AES-256-CBC encryption.  
However, there are two security concerns in any encrypted database that we have 
mitigated using the TPM:
\begin{itemize}
\item The database encryption key must be either stored or provided at startup, and 
it is susceptible to be stolen. To mitigate that we set use the motherboard TPM to 
generate and store a random key that is used to encrypt and decrypt arbitrary data 
using AES-256-CFB. Once setup, that key cannot be extracted from the TPM chip, and all
encryption and decryption happen inside of the chip without exposing the key. Then 
we generate a secrets file that is read and written through the TPM so we can access 
the database password but the file that stores the password can only be decrypted on 
the machine it was created. The secrets file can store multiple secrets using a single 
key inside of the TPM but uses IVs + XOR chaining of secrets to prevent 
statistics-based attacks. 

\item To query the database, it needs to be decrypted, and many pages can sit on the 
cache in plain text being vulnerable if an attacker could read the memory of this 
process. We use field encryption to mitigate this attack vector encrypting the QKD keys 
using AES-256-CFB with the TPM before being written to the database so ciphertext for 
the QKD Key is written in the Database and can be decrypted with TPM just in time if needed. 
It is important to note that when there is root, hypervisor, or physical level access, security 
cannot be guaranteed as an attacker could still use the TPM to decrypt the QKD keys but a memory 
dump without using the TPM to decrypt them does only expose the keys that are being actively relayed 
at the time of the memory dump. While these mechanisms protect stored secrets, they do not provide 
the runtime isolation guarantees of TEEs, and security cannot be guaranteed against sufficiently 
privileged attackers.
\end{itemize}

\subsection{Message formats \& error handling}

\begin{figure*}[t]
    \centering
    \includegraphics[width=0.75\linewidth]{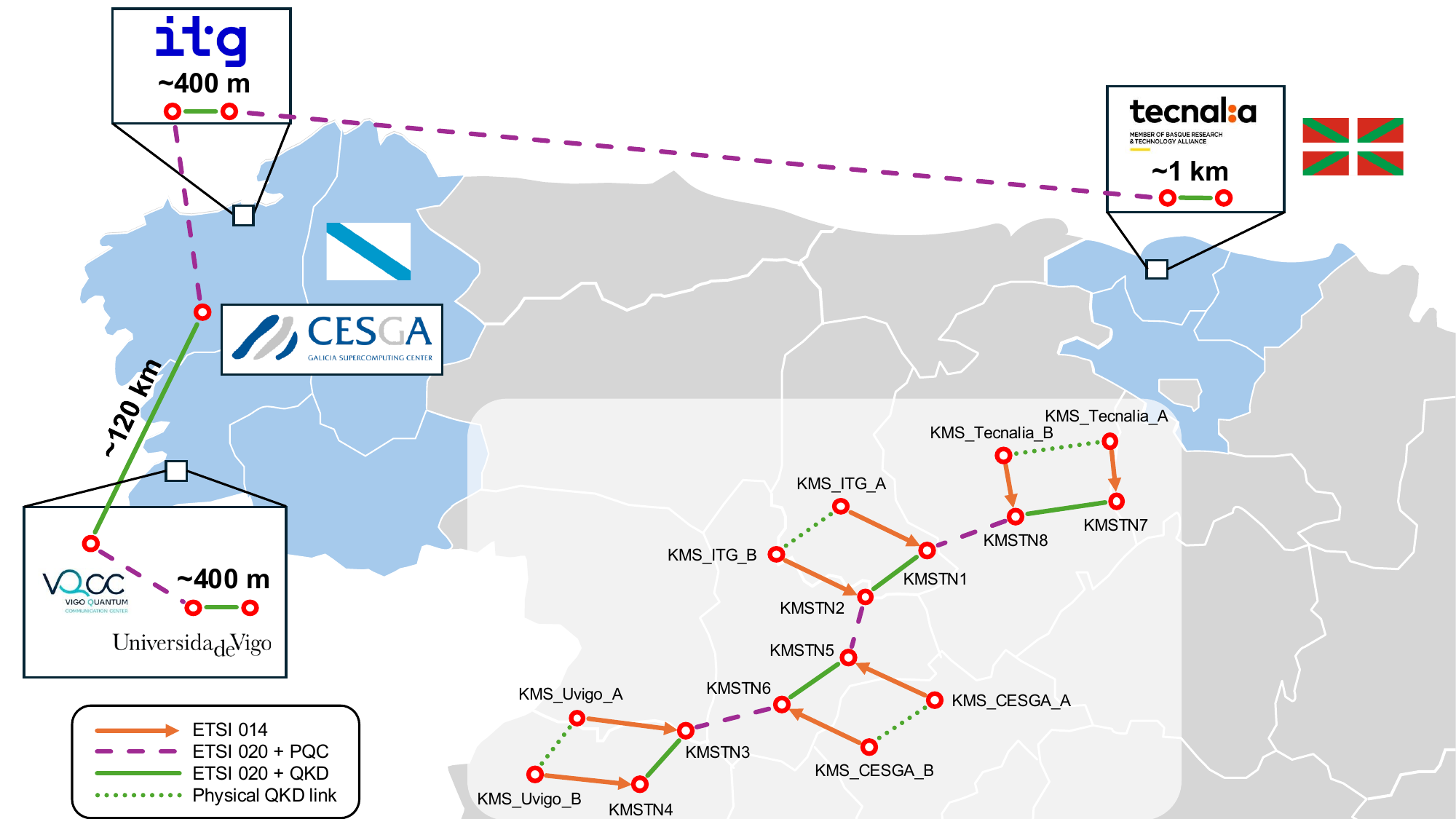}
    \caption{Map and network topology of our testbed, spanning the regions of Galicia and the Basque Country, 
    in Northern Spain. The solid green lines represent links which are secured via a physical QKD link, 
    through which keys can propagate via ETSI 020 using QKD + PQC encryption, and the dashed purple lines 
    show links where ETSI 020 interfaces were implemented using PQC encryption only. In the inset, the dotted 
    lines represent physical QKD links with vendor KMS, from which the KMSTN nodes extract QKD keys via ETSI 014 
    calls (orange arrows).}
    \label{fig:kmsgraph}
\end{figure*}

\subsubsection{Message Protocol Syntax}

The key routing protocol uses ETSI QKD GS 020 over HTTPS with mutual TLS as defined 
in the standard and, as explained above, a modification of these wherein the messages 
are further encrypted using AES-256-GCM. Not only the keys are encrypted, but the whole 
message is. Though the keys are certainly the most sensitive part, it is advantageous if 
the attacker cannot figure out anything about the message, including if the receiving
KMS is the final destination, or if it will be relayed. Each node on the route decrypts 
the message and encrypts it again with a different key and IV. 

The ciphered messages use a JSON format following this structure: \texttt{(iv, 
ciphertext, session, [SAE])}. The session acts as an ID for the ephemeral keys which 
can be derived by any means and exchanged with any desired frequency. In our
implementation, we are using a different key for each message, so they are fully 
ephemeral nonces. Our IDs represent either Kyber derived secrets, or the OTP of QKD 
keys and Kyber secrets.  When using QKD keys, the \textsc{sae\_id} that owns that key needs to 
be included, because there could be multiple QKD modes and they could serve multiple 
SAE each.  The IV is a unique public nonce generated for each AES-256-GCM encryption operation. 
It is transmitted together with the ciphertext and ensures nonce uniqueness across encryptions 
performed with the same session key. It does provide resistance against statistics-based attacks 
if the keys were to be reused. Currently, in our implementation, the keys are not reused, but this would 
enable us to improve the throughput by reusing keys without compromising the security unless AES-256-GCM 
was compromised.  

Once decrypted, the messages can contain any plain text, but in our case the payload
is only one of the messages defined by ETSI GS QKD 020.  The two most important are: 
\begin{itemize}
\item \texttt{ext\_key\_container}. This JSON message is used by one KMS to 
proactively share key to another KMS. It contains the keys, the \textsc{sae\_id} these keys 
belong to, and a list of SAEs the keys are being shared with. It has a URL for a 
callback to send an ack when the key is received by the destination. It is an 
asynchronous ACK that can leave at any later time. This is sent only after the key 
has been inserted in the destination KMS key store. It also contains mandatory and 
optional extension fields. The receiving KMS is allowed to ignore extensions in the
optional field, but if it receives a message with mandatory extensions that it has not 
implemented, it must reply with an error. We have not used any extensions. 

\item \texttt{ack\_containers}. This message is used by a KMS that has received 
a key through \texttt{ext-keys}. It confirms that the key was correctly received, 
and it is sent asynchronously to a callback URL that was provided in the original 
\texttt{ext-keys} message that shared the key.  \texttt{ack\_containers} is an 
array of acknowledgement containers, where each container follows this structure: 
1)  \texttt{key\_ids}: array of \texttt{key\_id\_container}; 2) \texttt{ack\_status}: 
enum of \texttt{[ relayed, voided, failed, key not present ]}; 3)
\texttt{initiator\_sae\_id}: the SAE that initiated the request message; 4)
\texttt{Optional message}: optional dictionary of extensions for future use. 
\end{itemize}

\subsubsection{Fallback behavior and error handling}

In the documents ETSI GS QKD 014 and ETSI GS QKD 020, there are a set of errors 
using HTTP status codes to add extra operational information. However, trying to 
recover from errors is not a requirement, only reporting them to the SAE. 
ETSI GS QKD 020 includes an acknowledgement system for \text{ext\_keys} and 
\texttt{void\_keys}. Our system sends the ACK once we can confirm that the key is 
in the key store of the destination KMSTN. If no ACK is returned, an error is shown. 
There are no attempts to retransmit the key, and it is up to the SAEs to handle the 
error. ETSI GS QKD 020 has a mechanism to void keys which we have implemented, 
although currently no void keys messages are emitted during normal operation. If one 
is received, it is processed and the key is voided. That could be part of more advanced
error handling. 
 

Besides, if there is no hardware TPM available to secure the KeyStore database key, 
we provide a fallback software TPM emulation inside of the docker container and produce
several warnings about the shortcomings of security implied by not having a real TPM. 
While almost all modern computers should have a hardware TPM, virtual machines do not
provide access to it unless it is configured in the hypervisor. If not even software 
TPM emulation is available, or if there is an issue with the TPM or for any other reason
the secrets file ca not be decrypted; we fall back to requesting the KeyStore key at every
startup. 

\begin{figure*}[t]
  \centering
  \includegraphics[width=\textwidth]{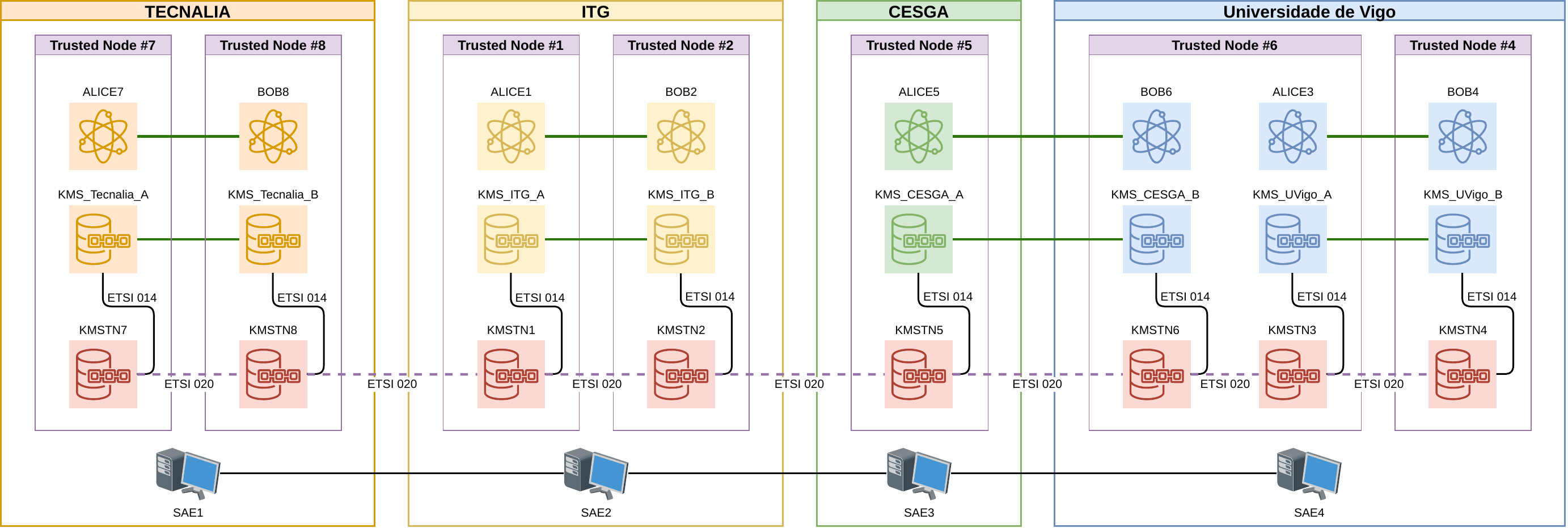}
  \caption{ Configuration for the testbed measurements.}
  \label{fig:trusted-nodes}
\end{figure*}

\subsection{Routing}

In the current prototype, routes are computed from a graph representation of the network topology 
using Dijkstra's algorithm at startup, with A*~\cite{Hart1968} available as a fallback. This approach 
provides automatic topology-aware path computation without requiring manual route configuration and enables 
optimal path selection when Dijkstra's algorithm is used. Although runtime dynamic routing is not currently 
implemented, the ETSI GS QKD 020 specification supports custom extensions that could be leveraged for 
this purpose. Furthermore, the system already includes mechanisms for handling ETSI GS QKD 020 extensions 
and tracking QKD node status, providing the foundations for future routing decisions based on network 
availability, key resources, or traffic load.

\section{Experimental setup}
\label{sec:testbed}

\subsection{Physical network}

Our QKD network consortium comprises four partners spanning two regions of Spain: Galicia 
and the Basque Country. They are the Galicia Supercomputing Center (CESGA) at Santiago de
Compostela, the atlanTTic Research Centre  (AtlanTTic) from the University of Vigo at Vigo, 
the Instituto tecnóloxico de Galicia (ITG) at A Coruña, and Tecnalia Research and Innovation
(TECNALIA) at Derio. The four partners have QKD equipment: CESGA a DV-QKD system from QTI 
(Quell-XR), atlanTTic a DV-QKD system from ID Quantique (Cerberis), ITG a CV-QKD system from
LuxQuanta (Nova-LQ), and TECNALIA a DV-system from ID Quantique (Clavis 3).

In Galicia, the QKD links used for this demonstration are located between CESGA and 
the Vigo Quantum Communications Center VQCC (Vigo), a second link in Vigo between
{VQCC and the central library of the University of Vigo, and a third in A Coruña. 
In the Basque country, a QKD link joins 204 building at Zamudio and 700 building at Derio. 
The long-distance links without QKD equipment between CESGA and ITG, and CESGA and 
TECNALIA, were integrated in the network using PQC protocols.  Our testbed measurements 
were carried out with a logical linear network topology (Fig.~\ref{fig:kmsgraph}), 
but are fully reconfigurable to, e.g., a star configuration through virtual PQC-link 
re-routing.

The testbed aims to exchange secure information across different Secure Application 
Entities (SAE) as shown in Fig.~\ref{fig:trusted-nodes}. The links between the various KMS and TN ensure 
the integrity and security of the key distribution process. Keys thus travel between the 
different nodes and systems, ensuring that the transmission adheres to PQC  encryption 
protocols and ETSI functional standards.  Notably, our QKD network does not include just 
the traditional architecture of quantum nodes and KMS, but also a crucial element: the 
Key Management System Trusted Nodes. KMSTNs form a layer above standard proprietary KMS, 
allowing the integration of diverse QKD "islands" scattered across entities, bridging 
different segments of the quantum network, and facilitating interoperability between nodes 
while maintaining PQC security. KMSTN present the following features:
\begin{itemize}
\item Connection of isolated "islands": the main advantage of KMSTN is that they enable
the connection of isolated segments. These segments may be operating with different 
quantum nodes but are not necessarily physically connected to each other. The KMSTN 
facilitates the creation of secure links between them through secure PQC encryption protocols,
acting as a trusted intermediary between the KMS of each entity. As shown in 
Fig.~\ref{fig:trusted-nodes}, the testbed network spans eight KMSTNs corresponding to four 
QKD links KMSTN1 and KMSTN2 (ITG), KMSTN3 and KMSTN4 (AtlanTTic-UVIGO), KMSTN5 and KMSTN6 
(CESGA), and  KMSTN7 and KMSTN8 (TECNALIA).

\item QKD between TNs: as the trusted nodes generate quantum keys, the KMSTN ensure that 
these keys can be transported between the different locations. In this context, the KMSTN 
allows a key generated, for example, in UVIGO, to be securely sent to TECNALIA, ensuring 
the integrity of the key during its transit through the intermediate nodes.

\item Compliance and interoperability: the KMSTNs implement protocols ETSI standards, 
allowing them to efficiently manage QKD and communication between different equipment 
and manufacturers.
\end{itemize}


\begin{table*}[t]
    \footnotesize
    \caption{\label{table:pairwise-keyrate} Pairwise keyrates among the nodes in the topology, average bps $\pm$ standard deviations. The figures were obtained with $100$ back-to-back key requests, without simultaneous or parallel requests by the \textsc{sae}s.}
    \begin{tabular}{c|cccccccc} \toprule
     & \textsc{kmstn1} & \textsc{kmstn2} & \textsc{kmstn3} & \textsc{kmstn4}
    & \textsc{kmstn5} & \textsc{kmstn6} & \textsc{kmstn7} & \textsc{kmstn8} \\ \midrule
    \textsc{1} & --- & $2625.7_{\color{blue} \pm 417.6}$  & $2472.4_{\color{blue} \pm 277.3}$ & $2457.0_{\color{blue} \pm 177.8}$
    & $2513.1_{\color{blue} \pm 244.9}$ & $2426.9_{\color{blue} \pm 247.8}$ & $2466.9_{\color{blue} \pm 245.1}$ & $2511.3_{\color{blue} \pm 260.7}$\\
    
    \textsc{2} & $2512.1_{\color{blue} \pm 503.2}$ & --- & $2408.2_{\color{blue} \pm 279.6}$ & $2341.2_{\color{blue} \pm 257.6}$
    & $2436.8_{\color{blue} \pm 268.2}$ & $2375.1_{\color{blue} \pm 233.3}$ & $2391.4_{\color{blue} \pm 383.7}$ & $2320.0_{\color{blue} \pm 367.2}$ \\

    \textsc{3} & $2314.5_{\color{blue} \pm 598.0}$ & $2255.6_{\color{blue} \pm 573.5}$ & --- & $2767.0_{\color{blue} \pm 810.5}$
    & $2454.5_{\color{blue} \pm 612.7}$ & $2355.0_{\color{blue} \pm 573.9}$ & $2349.2_{\color{blue} \pm 509.7}$ & $2248.91_{\color{blue} \pm 567.7}$ \\

    \textsc{4} & $1921.3_{\color{blue} \pm 322.1}$ & $1790.4_{\color{blue} \pm 358.2}$ & $1380.3_{\color{blue} \pm 217.9}$ & ---
    & $1767.4_{\color{blue} \pm 363.7}$ & $1904.1_{\color{blue} \pm 323.8}$ & $1834.2_{\color{blue} \pm 352.7}$ & $1895.3_{\color{blue} \pm 316.8}$ \\

    \textsc{5} & $368.6_{\color{blue} \pm 35.6}$ & $352.3_{\color{blue} \pm 42.3}$ & $370.9_{\color{blue} \pm 29.6}$ & $356.3_{\color{blue} \pm 46.6}$
    & --- & $381.0_{\color{blue} \pm 45.3}$ & $333.2_{\color{blue} \pm 48.9}$ & $346.1_{\color{blue} \pm 38.7}$\\

    \textsc{6} & $531.8_{\color{blue} \pm 66.4}$ & $541.6_{\color{blue} \pm 82.9}$ & $492.4_{\color{blue} \pm 72.0}$ & $500.5_{\color{blue} \pm 74.7}$
    & $489.3_{\color{blue} \pm 80.8}$ & --- & $469.1_{\color{blue} \pm 116.7}$ & $480.5_{\color{blue} \pm 93.94}$ \\

    \textsc{7} & $2362.2_{\color{blue} \pm 454.7}$ & $2360.3_{\color{blue} \pm 477.6}$ & $2419.6_{\color{blue} \pm 655.5}$ & $2316.1_{\color{blue} \pm 533.1}$ & $2402.7_{\color{blue} \pm 509.5}$ & $509.5_{\color{blue} \pm 526.0}$ & --- & $2276.8_{\color{blue} \pm 399.4}$\\

    \textsc{8} & $2045.9_{\color{blue} \pm 303.0}$ & $1998.3_{\color{blue} \pm 292.8}$ & $1980.0_{\color{blue} \pm 360.0}$ & $2000.4_{\color{blue} \pm 314.9}$ & $1843.1_{\color{blue} \pm 269.5}$ & $1978.4_{\color{blue} \pm 349.6}$ & $1881.6_{\color{blue} \pm 269.8}$ & --- \\ \bottomrule
    \end{tabular}
\end{table*}

\subsection{Software implementation}

The KMSTN server is distributed as a Docker container, which enables straightforward 
ans transparent deployment across heterogeneous environments. For the secure handling of 
local key material, it is recommended to use a hardware Trusted Platform Module (TPM). 
Specifically, the container will attempt to use \texttt{/dev/tpm0} or \texttt{/dev/tpmrm0}; 
if these are unavailable, it will fall back to a software TPM. In addition, when deploying 
inside a virtual machine, the hypervisor must be configured to expose the physical TPM to 
the guest; otherwise, hardware-backed protection will not be available. Moreover, when 
multiple KMSTN instances operate on the same physical host, distinct source directories must 
be used, as both the TPM and database directories are mapped directly from the host. Sharing 
these directories across instances can therefore lead to unreliable access to the key database.

The configuration of the Docker virtual machine relies on a set of structured
configuration files, which collectively define the behaviour of all node types 
participating in the network: real QKD nodes, trusted-node servers, post-quantum-secured 
servers, and Secure Application Entity clients (SAE files). These files describe both the
operational parameters of each component and the topology of the key-distribution environment.
For the present testbed, all configuration files must be present in every host to ensure 
that all KMSTN instances share a consistent and coherent view of the network, but this can 
be automated easily in systems at production scale.

In particular, the mode configuration files define the properties of physical QKD nodes,
including name, IP address, port, certificates, and the KME identifiers required by the
corresponding ETSI GS QKD 014 interfaces. Similarly,  the KMSTN configuration files define 
each server instance, including the aforementioned parameters, plus a port for PQC key 
exchange. For KMSTN, these files also specify the associated node via the 
\texttt{backing\_KMS\_conf} property. SAE configuration files, by contrast, define the 
IP addresses, identifiers, and certificates of SAE clients interacting with KMSTN servers. 
Once these settings are applied, each host intended to run a KMSTN instance must be 
reachable on both its configured service port and \texttt{pqc\_port} from outside its 
local network; otherwise, inter-KMSTN communication ---and consequently, key exchange--- 
will fail.

\begin{figure*}[t]
    \centering
    {\includegraphics[width=0.47\textwidth]{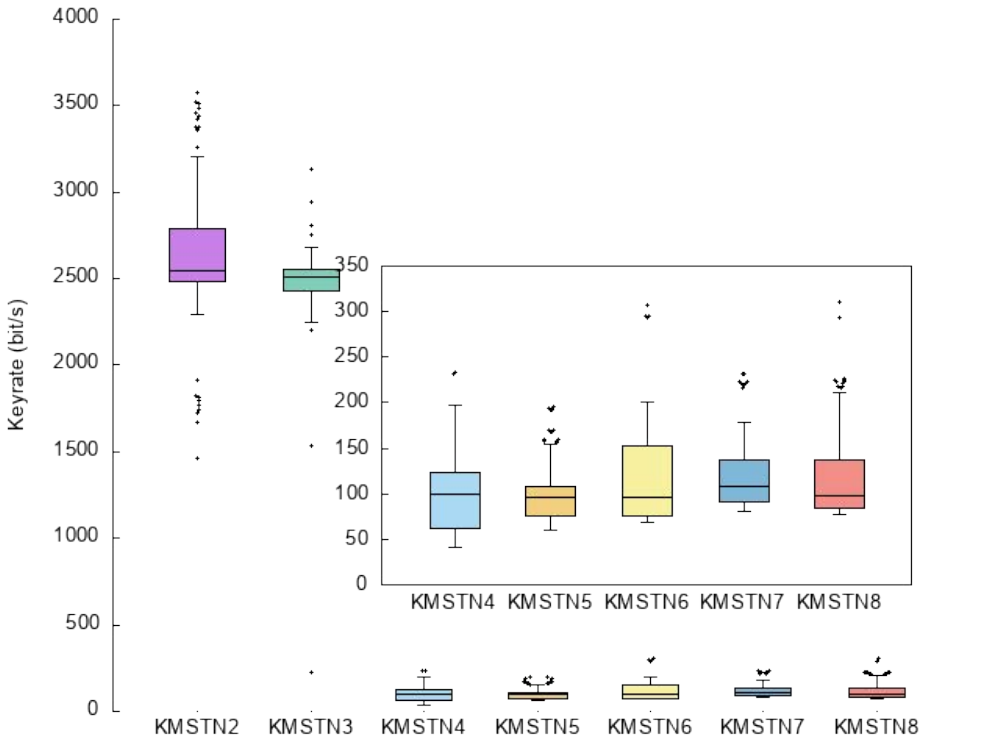}}
    {\includegraphics[width=0.47\textwidth]{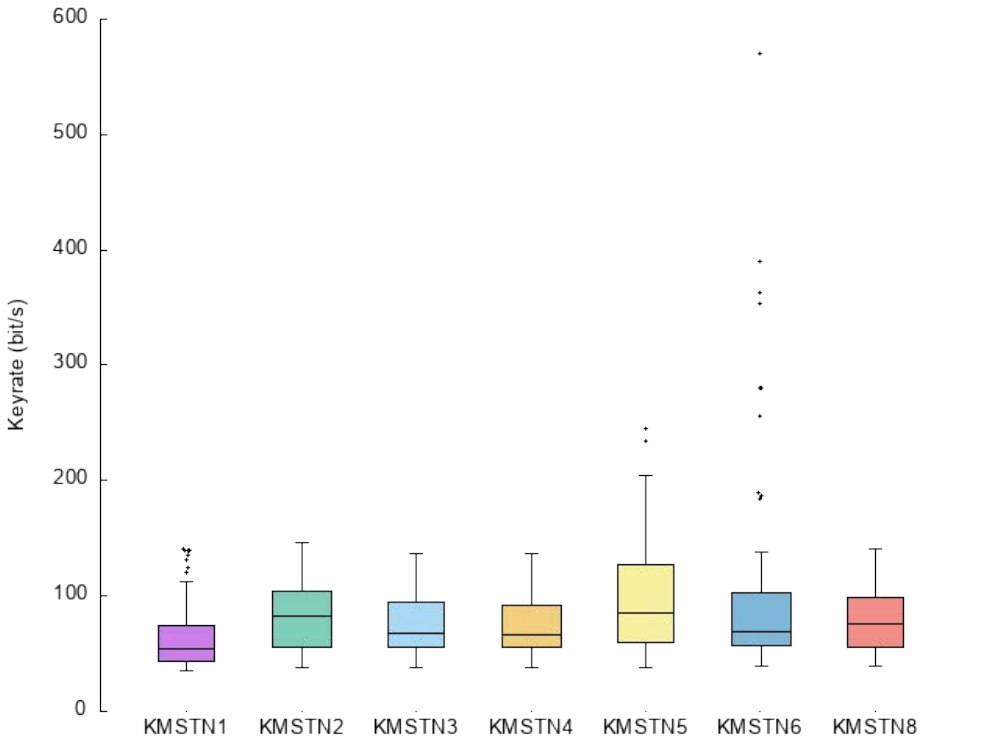}}
    \caption{Statistics of the keyrate for two nodes. KMSTN1 (left) and KMSTN7 (right)}
    \label{fig:keyrate-boxplots}
\end{figure*}

\section{Experimental Results and Performance Evaluation}
\label{sec:results}

\subsubsection{Setup}
We have conducted a series of experiments on our testbed to assess the fundamental 
performance metrics of the distributed KMS: keyrate, delay, and computational cost. 
The \emph{keyrate} is defined as the average number of bits per second obtained when continuously
requesting secret keys from the KMSTNs. \emph{Delay} is measured as the time elapsed 
from requesting  a keyblock (e.g., 128 bits) until the SAE gets the cryptographic material. 
And the consumption of resources in the proposed architecture and implementation is 
measured through the CPU and memory usage of the containers executing the KMSTN software, 
during the interval  where a test for keyrate or delay is being run. Our experimental 
measurements are not limited to sequential retrieval of the raw key material, on the contrary, 
concurrent key requests have also been tested controlled by a concurrency parameter, which 
is defined as the number of simultaneous key requests sent to a specific KMSTN. The concurrency
can thus be seen as a case where bursts of requests have to be processed by a KMSTN.
The following  paragraphs report the main results recorded throughout these tests.

\subsubsection{Keyrate}

Table~\ref{table:pairwise-keyrate} and Figure~\ref{fig:keyrate-boxplots} show the pairwise 
measurements obtained for the keyrate between the different KMSTNs. The average values of the 
keyrates are relatively low (around 2-3 kb/s, i.e., $10$ 256-bit keys per second, approximately,
but seem sufficient for supporting a moderate number of concurrent applications and users. There 
are noticeable differences among the nodes, however, that are explained by the diverse QKD 
technologies and link distances that comprise the system. This is particularly obvious for the 
nodes KMSTN5 and KMSTN6, which are separated by a $120$ km direct link and are capable of 
attaining only around $500$ b/s for the SKR. The distribution of the keyrates 
(Figure~\ref{fig:keyrate-boxplots})  clearly indicates that the values are skewed and also that 
a non-negligible fraction of the requests incur large deviations from the mean: there is a
high variability in the measured SKR that is ultimately related to the physical process of 
generating the key, but also to the difference between the production and consumption rates
of these keys, as will be explained later. This observation prompts to the need of
implementing some form of congestion control within the KMSTNs to throttle the global rate
of request that they receive.

\subsubsection{Delay}

The delay experienced in obtaining fresh keys by a SAE can be understood with the results 
shown in Table~\ref{table:delay-statistics} and in Figure~\ref{fig:delay-violinplot}. The 
median delay is consistently within the range $100$-$140$ ms, except for the two singular
nodes KMSTN5 and KMSTN6, wherein it raises substantially up to $700$ and $500$ ms, respectively.
This happens because, as mentioned, both nodes lie at the endpoints of a long-distance 
link having a distance at the edge of feasibility for current DV-QKD technology. Accordingly
to the expected reduction in the keyrate, the delay is also increased because this link
does not have a separate classical channel for key synchronization between the QKD nodes.
That is, the same optical fiber is used for the QKD key sifting process and for signalling 
the successful generation of keys, with most of the traffic and processing in the QKD nodes 
being dedicated to the key sifting and error correction phases of the QKD protocol. Clearly,
these processes introduce a high delay that could have a potential impact on end-user 
applications. The distribution of the delay (Figure~\ref{fig:delay-violinplot}) is clearly
skewed toward to right, meaning that very large delays (compared to the average) are observed 
with non-small probability. Again, the probability of extreme deviations is larger in the
two mutually farthest nodes, KMSTN5 and KMSTN6 (Figure~\ref{fig:delay-violinplot}, right panel).

\begin{table}[t]
    \small
    \caption{\label{table:delay-statistics} Statistics of delay for the key requests. Reported 
    values are the average delay (latency median, in ms), the standard deviation of the delay and 
    its $95$\%-th percentile (calculated with a moving window), and the median and $95\%$-th percentile 
    of the jitter. }
    \centering
    \begin{tabular}{c|cccc} \toprule
    \textsc{kmstn} & \textsc{jitter} & \textsc{95\%-th}  & \textsc{95\%-th} & \textsc{latency} \\ 
      \textsc{node} & \textsc{median} & \textsc{jitter}  & \textsc{roll\_std} & \textsc{median} \\
    \midrule
    1	& $5.5104$ & $42.079$ & $23.804$ & $102.661$ \\
    2	& $8.958$ & $53.33$ & $30.989$ & $106.473$ \\
    3	& $29.870$ & $78.944$ & $38.147$ & $110.638$ \\
    4	& $25.492$ & $91.555$ & $51.863$ & $140.890$ \\
    5	& $52.931$ & $359.525$ & $226.620$ & $704.282$ \\
    6	& $27.592$ & $724.589$ & $656.137$ & $496.933$ \\
    7	& $25.905$ & $67.390$ & $37.387$ & $114.441$ \\
    8	& $19.395$ & $59.999$ & $33.593$ & $131.394$ \\ \bottomrule
    \end{tabular}
\end{table}    
    
\begin{figure*}[t]
    \centering
    {\includegraphics[width=0.47\textwidth]{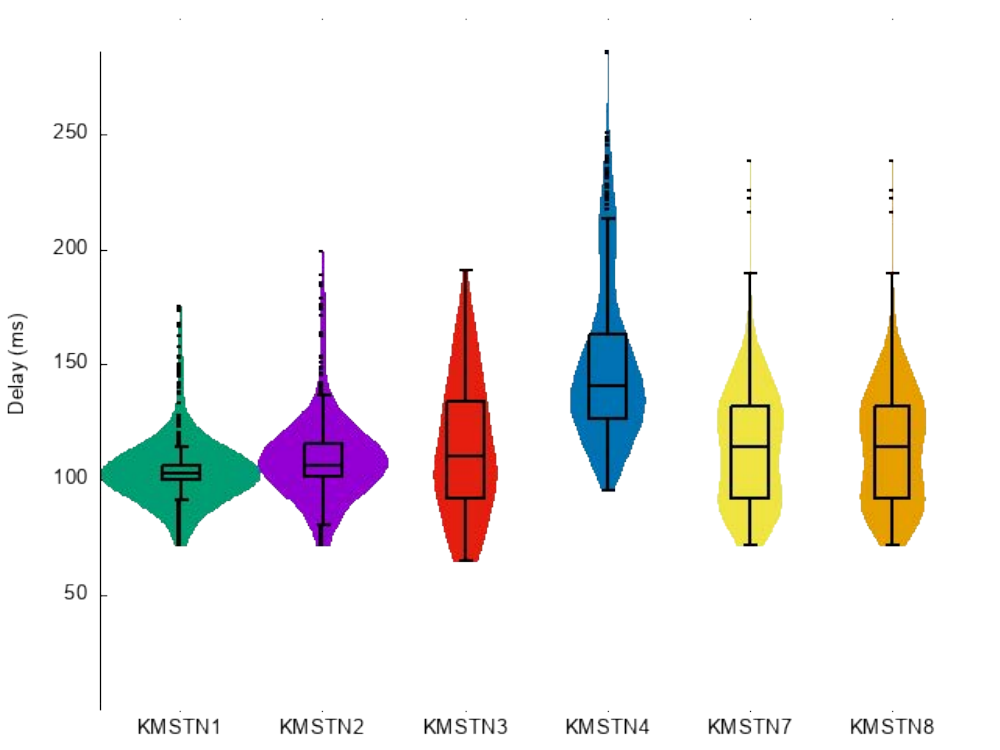}}
    {\includegraphics[width=0.47\textwidth]{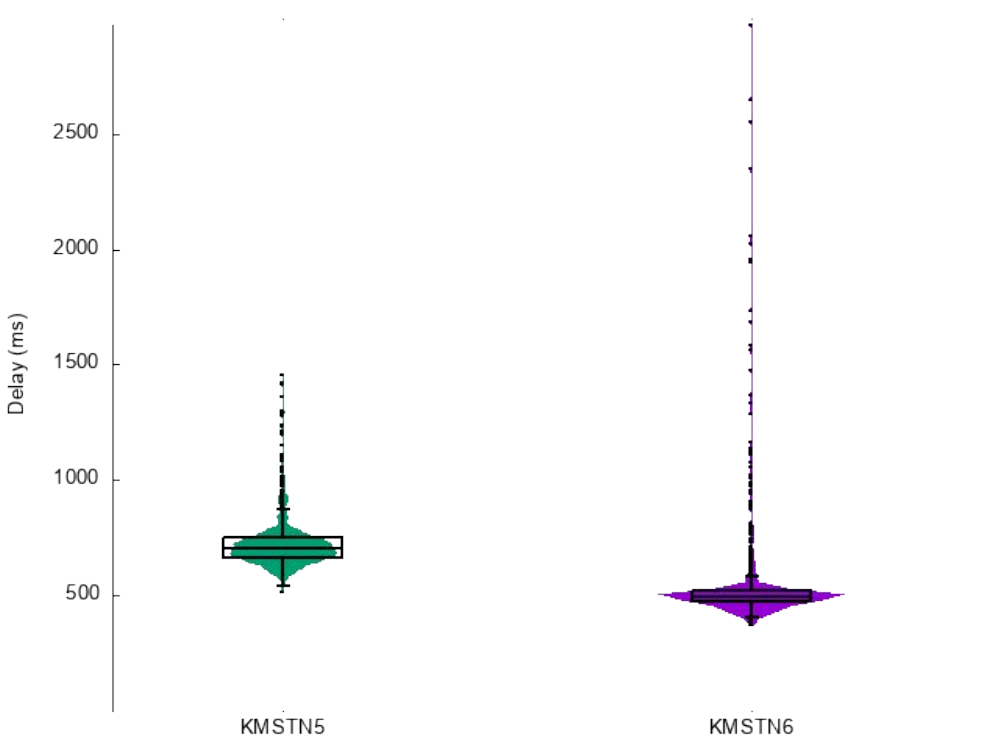}}
    \caption{Statistics of delay for several nodes.}
    \label{fig:delay-violinplot} 
\end{figure*}
\begin{figure*}[t]
    \centering
    {\includegraphics[width=0.95\textwidth]{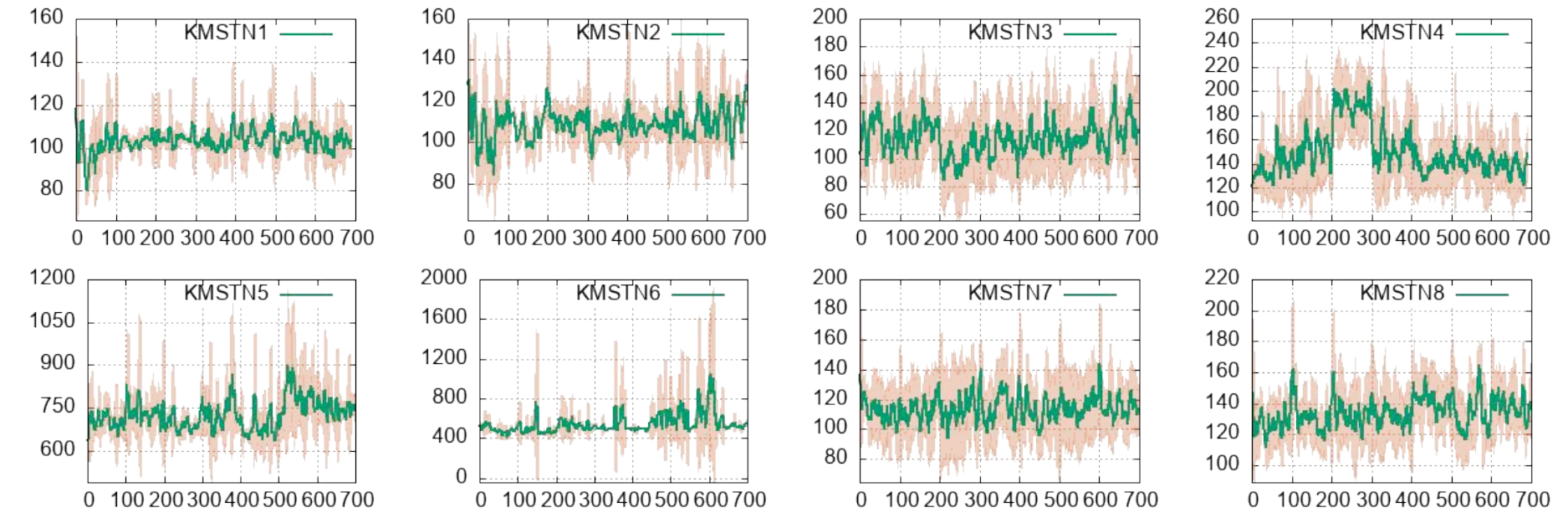}}
    \caption{Latency of key requests in ms. Values are averaged over a $4$ sample window, and the 
    standard deviation is plotted as a shaded area.}
    \label{fig:rolling-latency} 
\end{figure*}

Figure~\ref{fig:rolling-latency} illustrates the variability in time of the latency
for obtaining the keys. Average values $\pm$ one standard deviation (shaded area)
are plotted for each if the $8$ nodes, after smoothing the individual measurements with a 
sliding window of $4$ points. The traces confirm that, in a short-time scale, the 
latency values show a significant variability over time, which might have a potential
impact on application performance.

Figure~\ref{fig:latency-vs-time-a} depicts the latency measured in an experiment where $100$ 
concurrent requests (the vertical dots at each time epoch) are sent to a specific node. A
fraction of the requests fail (red dots in the subfigures), as a consequence of the congestion 
at the nodes, causing the internal KMS buffer of keys to empty and the KMS to respond
with an error message to the API call. The dispersion in the latency of concurrent requests is
clearly visible in the vertical alignment of the measurements, and confirms the high 
variability in the round-trip time from the issue of a request until its matching response 
arrives to the SAE. The measurements forming a diagonal line are simply delayed responses
caused by the time employed by the QKD nodes in generating new key while incoming requests
are being received.

\begin{figure*}[t]
    \centering

    \subfigure[Latency of 100 concurrent key requests vs. time. 
    Each data point is a measure of delay, and the vertical slices correspond to queries made 
    to a specific KMSTN. \label{fig:latency-vs-time-a}]{%
    \includegraphics[width=0.9\textwidth]{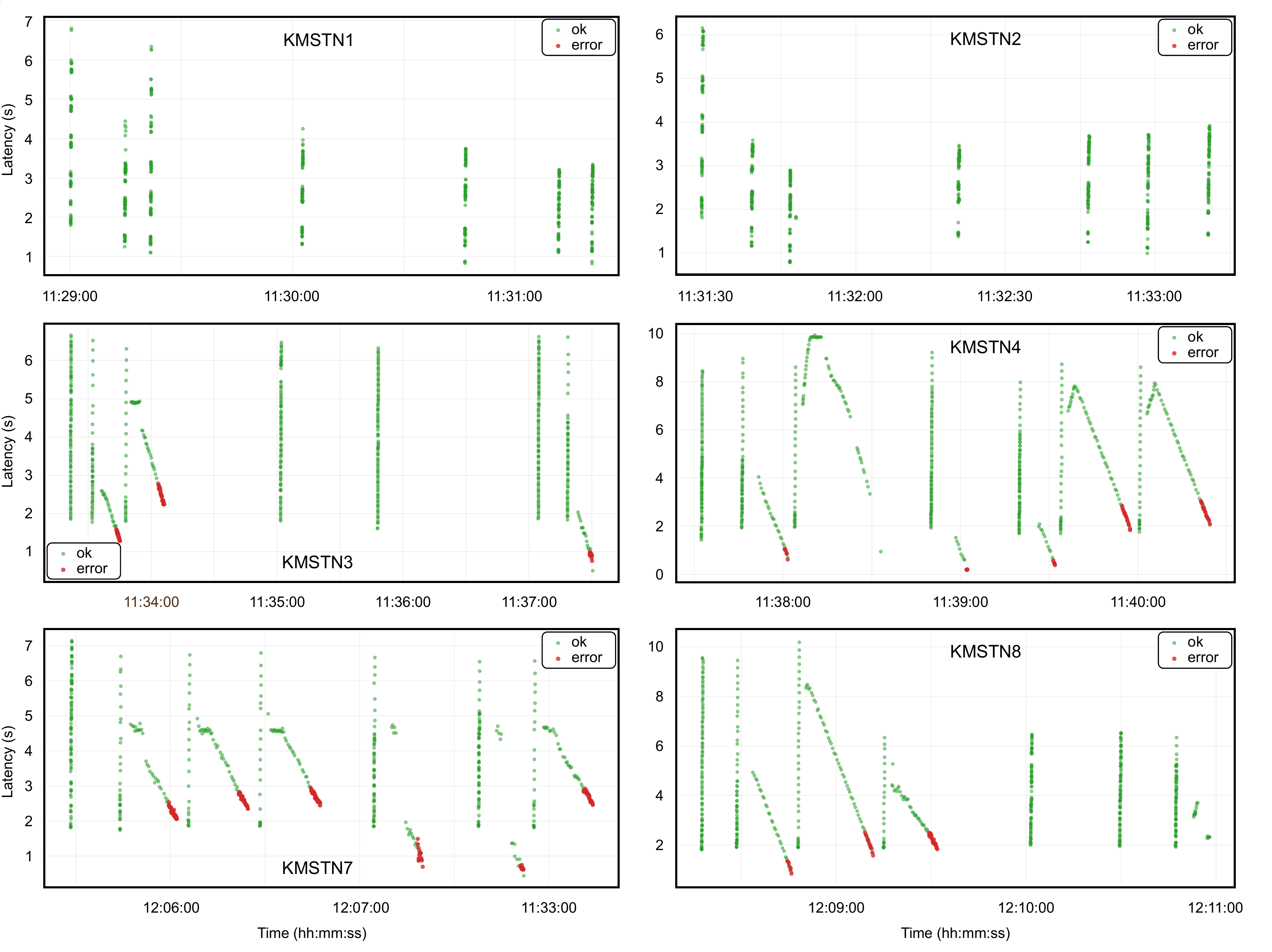}}
    \subfigure[Latency of 10 concurrent key requests vs. time. \label{fig:latency-vs-time-b}]{%
    \includegraphics[width=0.45\textwidth]{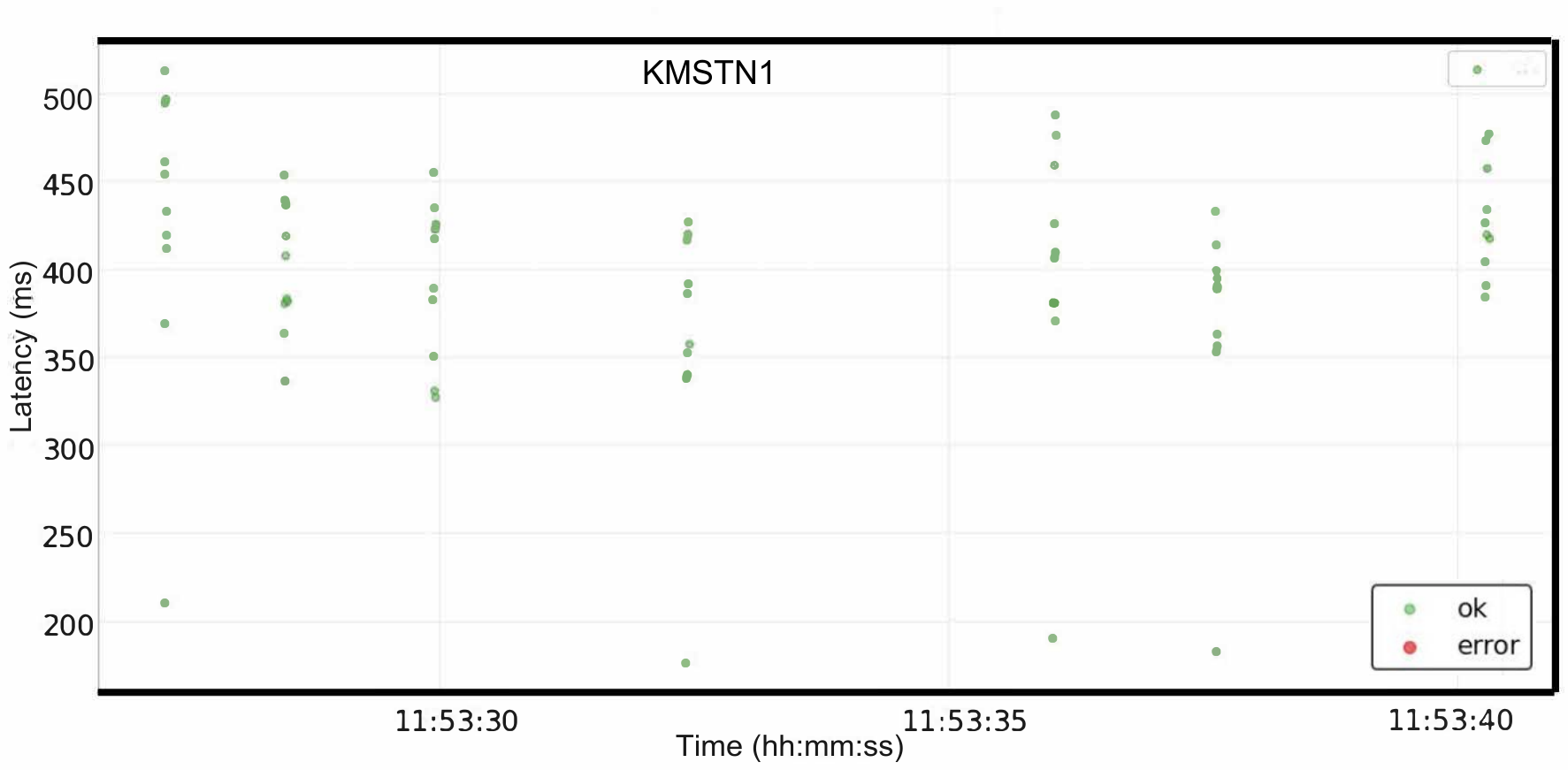}
    \includegraphics[width=0.45\textwidth]{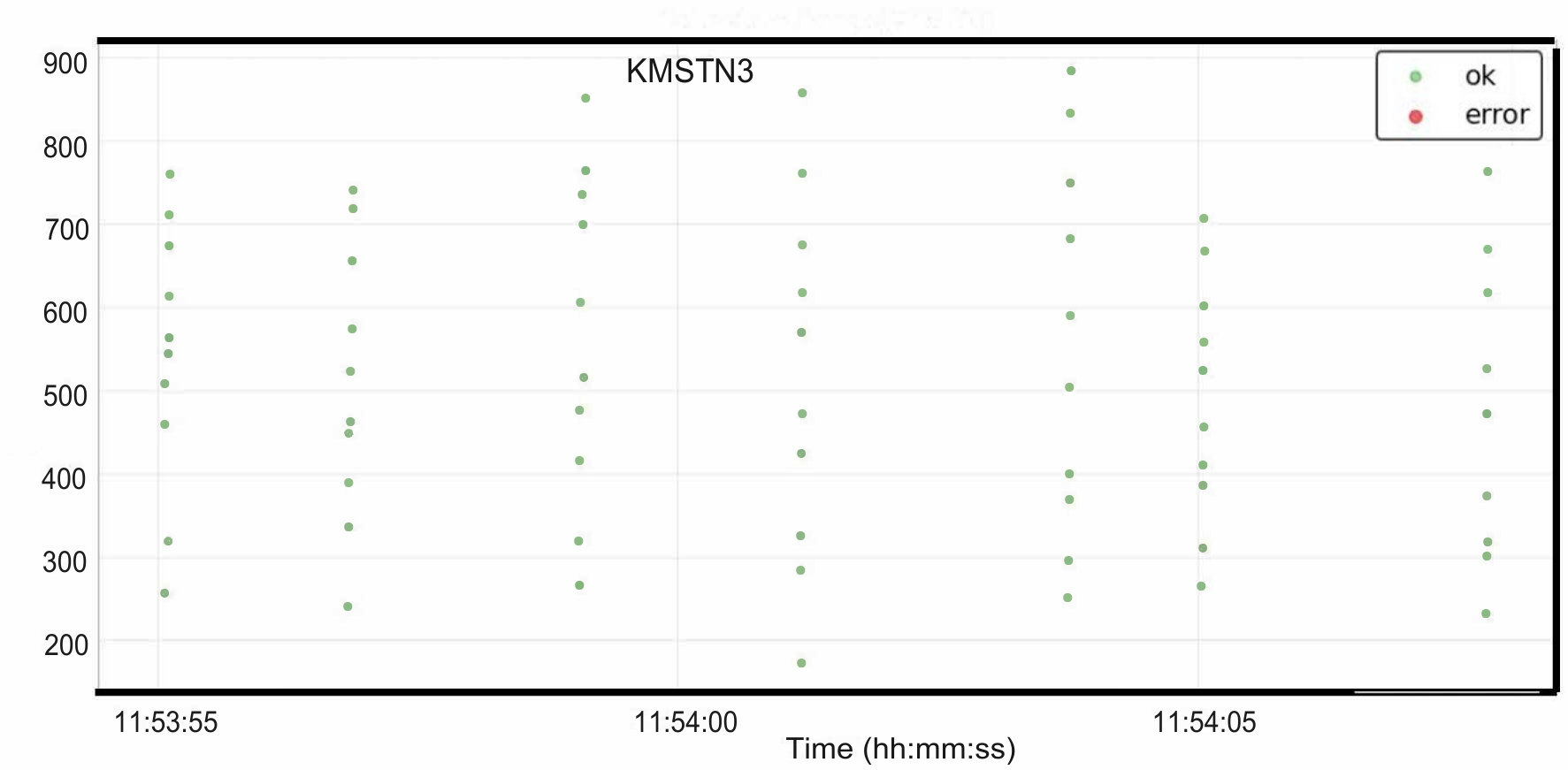}}
    \caption{Latency for key delivery across KMSTNs and different number of concurrent requests.}
    \label{fig:latency-vs-time}
\end{figure*}

\subsubsection{Correlations}
We also measured (Figure~\ref{fig:correlations}}) the empirical correlations 
between the metrics captured across the experiments (error rate, average SKR, 
$95\%$-th percentile of SKR) and the variables that define the 
pattern of the key requests: 1) number of keys requested, 2) total requests, and 3) concurrency (the number of
simultaneous key requests, simulating concurrent users in the network). The 
values in the correlation matrices (for two nodes located at different QKD network
segments, KMSTN1 and KMSTN8) are generally not significant, with
absolute correlation strength below $0.5$, except for a 
statistically measurable dependence between the mean key rate and its 95\% percentile 
value (\texttt{KR mean} and \texttt{KR p95}), and between the number of 
requests (\texttt{numkeys}) and the error rate. The first case corroborates the
results in Figs.~\ref{fig:keyrate-boxplots}, the statistical distribution of the
key rate is highly concentrate around its mean value, with only a few outlier points
of extreme high value. For the second case, the high correlation ($\approx 0.6$ to $0.8$)
is indicative that a high volume of key requests in short periods of time are
likely to deplete the buffer of the KMSTN and trigger a failure. We discuss 
a possible solution to these failures in the next paragraph.

\begin{figure*}[t]
    \centering
    \subfigure[KMSTN1]{%
    \includegraphics[width=0.42\textwidth]{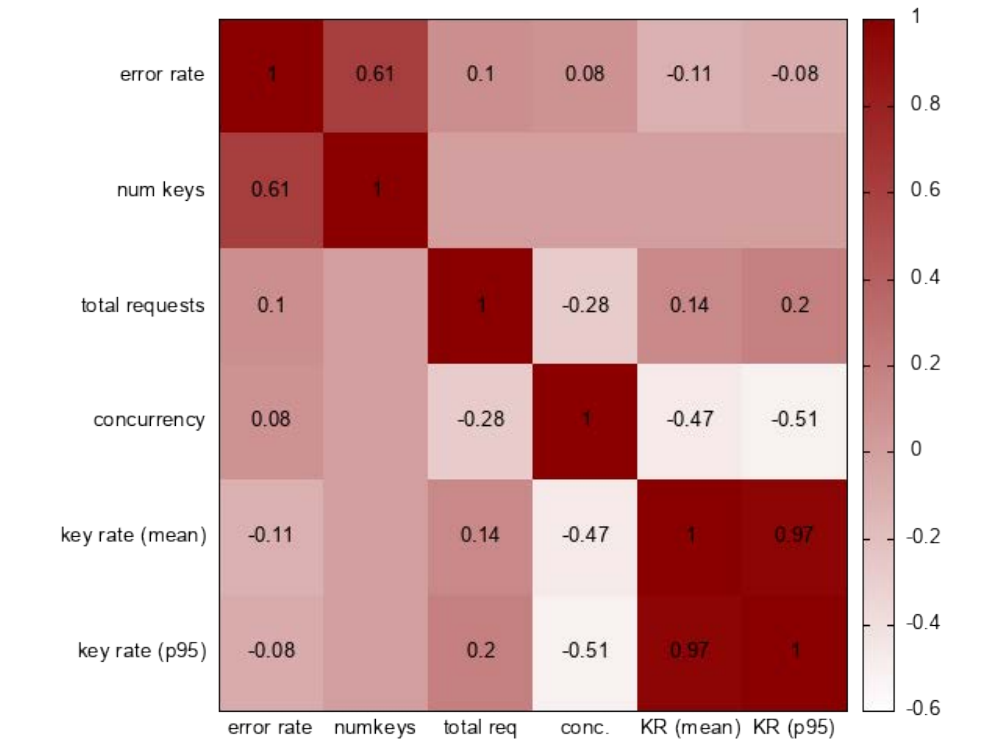}}
    \subfigure[KMSTN8]{%
    \includegraphics[width=0.42\textwidth]{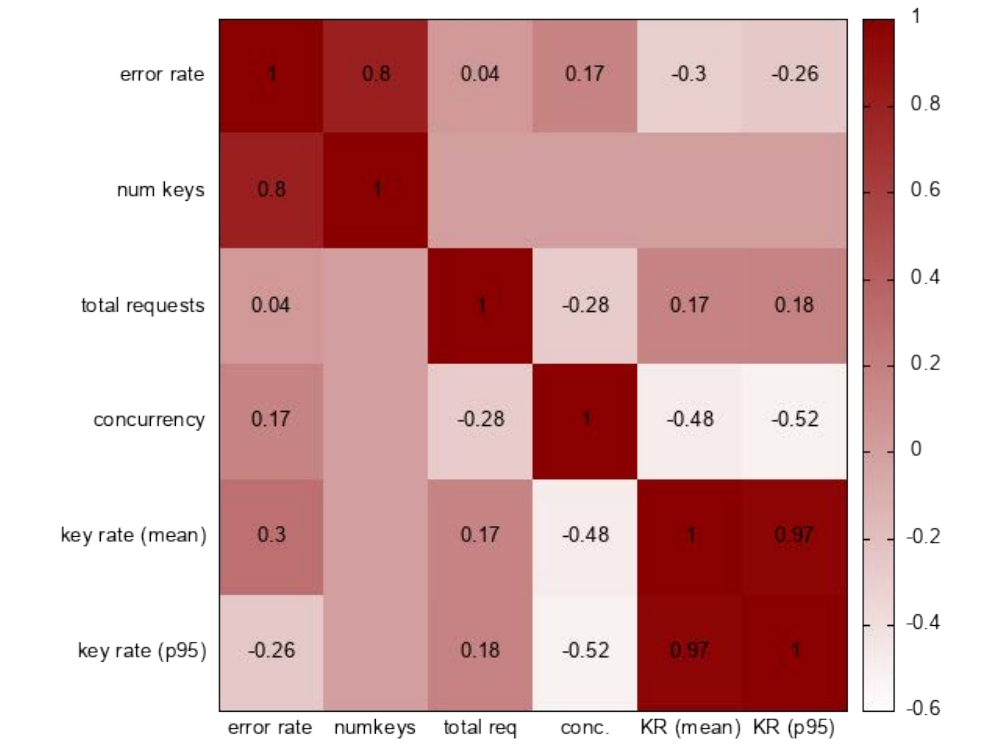}}
    \caption{\label{fig:correlations} Pairwise correlation matrix between the metrics collected 
    in the experiments for two KMSTN nodes: (a) KMSTN1, (b) KMSTN8.}
\end{figure*}

\subsubsection{Scalability}

The number of physical connections and QKD devices in our testbed was limited due to the high 
cost of acquisition and fiber deployment. For this reason, to assess how key performance metrics 
potentially scale in larger networks and validate our design choices and operational principles 
to larger deployments, we adopted the following approach on the existing infrastructure.

We measured performance on the 8 KMSTNs with a variable load of key requests, 
simulating a network with moderate to large number of users, where an user is any 
arbitrary application or process requesting keys ---hence the number of users is not 
necessarily equal to the number of SAEs. In Figure~\ref{fig:latency-vs-time-b},
the results for two representative nodes are shown, corresponding to 10 
simultaneous requests issued to each node. Each group of requests is repeated 
several times separated by an idle interval. Compared to Fig.~\ref{fig:latency-vs-time-a}
the results confirm that the delay until the users obtain the application keys grows up with 
the peak value of the number of requests, with the cause of the bottleneck attributed to two factors: 
inter-node key encapsulation and propagation, and servicing of the key requests in a sequential 
manner. Parallelizing the incoming requests to the KMSTNs can alleviate the latter 
problem, and does not require architectural changes.  However, the key forwarding 
time is clearly proportional to the path length. In large networks, this suggests 
that the virtual topology formed by the KMSTNs should have a low diameter.

In turn, a high load on a KMSTN can potentially exhaust the buffer of keys read out 
from the QKD links, as Fig.~\ref{fig:latency-vs-time}  show  (red dots). A solution to these congestion 
events is to divert the requests arriving to a congested node toward other KMSTNs with less load, since all 
the KMSTNs are functionally equivalent. This would need a central controller periodically monitoring
the state of buffers across the network, and deciding which node to use for 
delivering the new requests. Hence, scalability in large networks requires an
additional key allocation layer that manages all the KMSTNs as a distributed and 
homogeneous substrate for generating the keys, regardless of the origin of the requests.

\section{Discussion}
\label{sec:discussion}

This work demonstrates the feasibility of interconnecting several regional QKD 
infrastructures through a standards-compliant overlay that enables end-to-end key delivery 
across heterogeneous domains. Beyond validating the technical integration, the 
implementation reveals several architectural properties that shape how inter-regional 
QKD services can be designed and operated.

\paragraph{Separation between quantum generation and service delivery}
The architecture decouples physical QKD devices from the service-facing key delivery layer. 
QKD systems remain locally integrated with their vendor KMSs, while KMSTN instances abstract 
key material as a routable resource across domains. This enables inter-regional key delivery 
without exposing applications to device-level heterogeneity or requiring a continuous 
quantum path.

\paragraph{Standards-based federation}
Interoperability is enforced through strict adherence to ETSI GS QKD 014 and 020. Rather 
than modifying proprietary KMS implementations, the overlay federates existing key pools 
into a unified inter-regional service. The result is a standards-aligned integration 
layer capable of bridging heterogeneous domains with minimal intrusion.

\paragraph{Portable and reproducible deployment model}
Containerisation and configuration-driven topology management ensure that the system 
can be instantiated consistently across distributed infrastructures. This portability 
supports controlled experimentation and scalable deployment in environments characterised 
by administrative separation and infrastructure diversity.

\paragraph{Security role of QKD in the hybrid architecture}
QKD and PQC provide complementary security properties. PQC enables quantum-resistant connectivity where QKD links are unavailable, while QKD contributes secret entropy whose security is independent of computational hardness assumptions. On QKD-enabled links, combining QKD and ML-KEM material provides defence in depth, since compromise of the PQC mechanism alone does not reveal the hop-protection key.

The architecture therefore does not claim information-theoretic security across arbitrary hybrid paths. Instead, QKD reduces reliance on a single computational assumption, preserves information-theoretic key generation within regional domains, and supports their federation without requiring continuous QKD connectivity.

\subsection{Interpretation of performance behaviour}

The keyrate measurements under increasing concurrency, as shown in Figure~\ref{fig:keyrate-boxplots} for KMSTN1 and KMSTN7, indicate that throughput is jointly constrained by the effective supply rate of the underlying QKD links and by the per-hop processing overhead introduced by the overlay. As the number of simultaneous requests grows, the system approaches a regime in which forwarding, buffering, and synchronisation effects at intermediate nodes become visible, rather than request generation alone determining performance.

The large-scale request experiment illustrated in Figure~\ref{fig:delay-violinplot} further highlights how sustained demand interacts with the availability of buffered key material. Unlike a point-to-point QKD link, where keys are continuously generated and accumulated in local buffers, inter-node key propagation in the overlay is triggered on demand. This architectural difference introduces additional latency components associated with route traversal, per-hop encryption and storage operations, and coordination between KMS instances.

The on-demand propagation model makes end-to-end latency a particularly relevant metric, as it directly reflects the time required to make fresh key material available across multiple hops, in contrast to the steady-state buffering behaviour of standalone QKD links.

\subsection{Limitations}

Despite the feasibility demonstrated in the testbed, several structural and operational limitations remain.

\textbf{Static routing behaviour:} Route selection is computed at start-up from a 
predefined topology graph and does not adapt dynamically to link conditions, key pool levels, 
or load fluctuations. Consequently, the system operates with static optimal paths rather 
than reacting in real time to network state variations.

\textbf{Configuration scalability:} Overlay connectivity depends on consistent 
configuration across participating nodes. While manageable at pilot scale, this approach 
may become increasingly complex as the number of domains, sites, and administrative 
boundaries grows.

\textbf{Trusted-node exposure:} The architecture follows a trusted-node relaying paradigm 
in which intermediate nodes process and re-encrypt key material at each hop. As a result, 
these nodes remain part of the trust boundary and constitute potential exposure points.

\textbf{Authentication:} Node authentication currently relies on conventional 
certificate-based mechanisms integrated into the TLS stack. Although this provides 
transport-level authenticity and integrity, the integration of quantum-safe or 
QKD-derived authentication primitives has not yet been implemented and remains an open 
research and engineering challenge.

\section{Conclusion}
\label{sec:conclusion}

In this work, we presented the design, implementation, and empirical validation of a federated Key Distribution Service capable of bridging geographically dispersed QKD islands over a 500 km production-grade testbed spanning Galicia and the Basque Country in Spain. Our primary contribution is the development of a robust hybrid QKD-PQC architecture ---strictly adhering to ETSI GS QKD 014 and 020 standards--- that securely extends key routing across non-QKD classical WAN segments. Kyber was integrated into TLS to attain post-quantum security
on the horizontal ETSI QKD GS 020 messages, as a replacement of the classical Diffie-Hellman handshake. Furthermore, the use of hardware Trusted Platform Modules hardens the multi-hop relay process by encrypting the local store databases to minimize the attack surface.

The significance of this hybrid key transport paradigm lies in its ability to provide a unified key-delivery service across discontinuous QKD infrastructures, without requiring a contiguous physical QKD path between communicating domains. By decoupling the key distribution service from the need for continuous dark-fibre deployments, the architecture avoids geographical bottlenecks and supports continental-scale federation for initiatives such as EuroQCI. Furthermore, the system demonstrates a high degree of deployment readiness for integration into carrier and operator infrastructures, as its containerised Key Management System is compatible with modern orchestration frameworks.

Future work should focus on introducing dynamic control mechanisms capable of adapting  routing decisions to real-time key availability and link status. Strengthening the authentication layer with quantum-safe or QKD-assisted approaches is also essential to align all security components with long-term post-quantum requirements. Systematic benchmarking  of end-to-end latency and resource consumption under realistic workloads remains necessary, particularly to characterise the propagation delay introduced by on-demand multi-hop key delivery. Closer integration with consumption layers such as IPsec or TLS endpoints will allow evaluation under protocol-driven traffic and further consolidate the architecture as an operational inter-regional QKD key delivery service.

\bibliographystyle{ieeetr}
\bibliography{biblio}

\end{document}